\documentclass[iop,revtex4]{emulateapj}

\usepackage{color,commath}

\slugcomment {}

\shorttitle{NEOWISE proper motion survey}
\shortauthors{Schneider et al.}

\begin{document}

\title{A Proper Motion Survey Using the First Sky Pass of NEOWISE-Reactivation Data}

\author{Adam C. Schneider\altaffilmark{a}, Jennifer Greco\altaffilmark{a}, Michael C. Cushing\altaffilmark{a}, J. Davy Kirkpatrick\altaffilmark{b}, Amy Mainzer\altaffilmark{c}, Christopher R. Gelino\altaffilmark{b,d}, Sergio B. Fajardo-Acosta\altaffilmark{b}, \& James Bauer\altaffilmark{c}}  

\altaffiltext{a}{Department of Physics and Astronomy, University of Toledo, 2801 W. Bancroft St., Toledo, OH 43606, USA; Adam.Schneider@Utoledo.edu}
\altaffiltext{b}{Infrared Processing and Analysis Center, MS 100-22, California Institute of Technology, Pasadena, CA 91125, USA}
\altaffiltext{c}{Jet Propulsion Laboratory, California Institute of Technology, Pasadena, CA 91125, USA}
\altaffiltext{d}{NASA Exoplanet Science Institute, Mail Code 100-22, California Institute of Technology, 770 South Wilson Ave, Pasadena, CA 91125, USA}

\begin{abstract}
The {\it Wide-field Infrared Survey Explorer (WISE)} was reactivated in December of 2013 (NEOWISE) to search for potentially hazardous near-Earth objects.  We have conducted a survey using the first sky pass of NEOWISE data and the AllWISE catalog to identify nearby stars and brown dwarfs with large proper motions ($\mu_{\rm total}$ $\gtrsim$ 250 mas yr$^{-1}$).  A total of 20,548 high proper motion objects were identified, 1,006 of which are new discoveries.  This survey has uncovered a significantly larger sample of fainter objects (W2 $\gtrsim$13 mag) than the previous {\it WISE} motion surveys of \cite{luh14a} and \cite{kirk14}.  Many of these objects are predicted to be new L and T dwarfs based on near- and mid-infrared colors.  Using estimated spectral types along with distance estimates, we have identified several objects likely belonging to the nearby Solar neighborhood (d $<$ 25 pc).  We have followed up 19 of these new discoveries with near-infrared or optical spectroscopy, focusing on potentially nearby objects, objects with the latest predicted spectral types, and potential late-type subdwarfs.  This subset includes 6 M dwarfs, 5 of which are likely subdwarfs, as well as 8 L dwarfs and 5 T dwarfs, many of which have blue near-infrared colors.  As an additional supplement, we provide 2MASS and AllWISE positions and photometry for every object found in our search, as well as 2MASS/AllWISE calculated proper motions.             

\end{abstract}


\section{Introduction}
Nearby stars and brown dwarfs serve as benchmarks for many vital areas of astrophysics, both as individual objects and as an ensemble.  As individual objects, they are particularly attractive as astrophysical laboratories because they are the brightest examples of their spectral type, and are therefore optimal targets for detailed studies of a given class.  Nearby brown dwarfs also provide the best examples with which to study cold, exoplanet-like atmospheres across a variety of physical parameters (e.g., surface gravity, metallicity), thus offering critical checks of theory.  As a population, nearby low-mass objects probe the efficiency (or lack thereof) of star formation at low masses, and as a result provide detailed information on the shape and cut-off of the initial mass function in a regime ($<$ 30 Jupiter masses) that is difficult to study in sites of active star formation (e.g., \citealt{kirk12}).  Cataloguing the nearest Solar neighbors, however, is not a straightforward procedure.  Indeed, recent discoveries have demonstrated that some of our closest neighbors have been lurking unseen because of their low temperatures and luminosities (e.g., WISE J052126.29$+$102528.4 ($\sim$5 pc), \citealt{bih13}; WISE J104915.57$-$531906.1AB ($\sim$2 pc), \citealt{luh13}; WISEA J154045.67$-$510139.3 ($\sim$5.9 pc), \citealt{kirk14}; WISE J085510.83$-$071442.5 ($\sim$2 pc), \citealt{luh14b}; and WISE J072003.20-084651.2 ($\sim$7 pc), \citealt{scholz14}).

Most searches for nearby very-low-mass stars and brown dwarfs have used red optical through mid-infrared colors as the main selection criterion due to the shift in the peak wavelength of the Planck function with decreasing effective temperature (e.g.\ {\citealt{kirk99}, \citealt{leg00}, \citealt{kirk11}). While such searches are geared toward finding objects with normal gravities and solar-like metallicities, they are generally biased against uncovering objects with unusual characteristics. Kinematic searches, on the other hand, avoid such a bias by using proper motion alone as a judge of distance. By identifying objects with large proper motions, unusual brown dwarfs overlooked by previous surveys can be identified (e.g., \citealt{met08}, \citealt{dea09}, \citealt{shep09}, \citealt{art10}, \citealt{kirk10}, \citealt{dea11},  \citealt{liu11}, \citealt{giz11}, Scholz et al.\ 2011, 2012, 2014).

Multi-epoch data from the {\it Wide-field Infrared Survey Explorer} {\it (WISE)} have enabled the first all-sky motion searches using solely mid-infrared wavelengths, which allows for the straightforward identification of low-mass stars and brown dwarfs in a wavelength region where they emit their peak flux.  The motion survey of \cite{luh14a} and the AllWISE motion survey \citep{kirk14} used data from the primary {\it WISE} mission to uncover thousands of new objects with significant proper motions, including the aforementioned WISE J104915.57$-$531906.1AB, WISE J085510.83$-$071442.5, and WISEA J154045.67$-$510139.3, along with a wealth of previously unknown late-type subdwarfs (\citealt{kirk14}, \citealt{luh14c}).  Most of these discoveries were identified using a six month time baseline between {\it WISE} epochs ($\sim$20\% of the sky was covered with an additional third epoch, resulting in a time baseline of one year).  Despite the successes of the Luhman and Kirkpatrick et al.\ studies, each survey missed objects that the other one found, due to their different candidate selection procedures, suggesting that there are likely more nearby objects to be discovered.  Of the 3525 and 762 discoveries in \citealt{kirk14} and \citealt{luh14a}, respectively, only 321 were common to both surveys.      

{\it WISE} was reactivated in December of 2013 to search for potentially hazardous near Earth objects (NEOWISE; \citealt{main14}).  We have completed a survey whereby we used the individual detections from the first NEOWISE pass of the sky in combination with the AllWISE source catalog \citep{cut13} to identify previously overlooked stars and brown dwarfs with large proper motions.  Our goals are to 1) identify late-type subdwarf candidates to further map the existence and extent of the putative subdwarf gap \citep{kirk14} in order to place constraints on brown dwarf cooling theory, 2) identify overlooked nearby stars and brown dwarfs, which can have a significant impact on investigations of the initial mass function of the local population, and 3) identify brown dwarfs with unusual characteristics (e.g., binaries).  In Section 2, we describe the search strategy for the  NEOWISE proper motion survey, while the results are presented in Section 3.  In Section 4 we describe the follow-up spectroscopic observations and analysis of a subset of discoveries from this effort.     

\section{Identifying Objects with High Proper Motions}
The NEOWISE reactivation mission was carried out using the W1 (3.4 $\mu$m) and W2 (4.6 $\mu$m) passbands of the {\it WISE} telescope.  Because the intent of the NEOWISE observations is the identification of near-Earth objects, the images are not co-added like the previous epochs of {\it WISE} data.  However, the detections from each individual {\it WISE} frame are collected into a single catalog. Since the NEOWISE images are not co-added, the first step in identifying high proper motion objects is to construct a source catalog from the NEOWISE Single Exposure Source Table.  

\begin{figure*}
\plotone{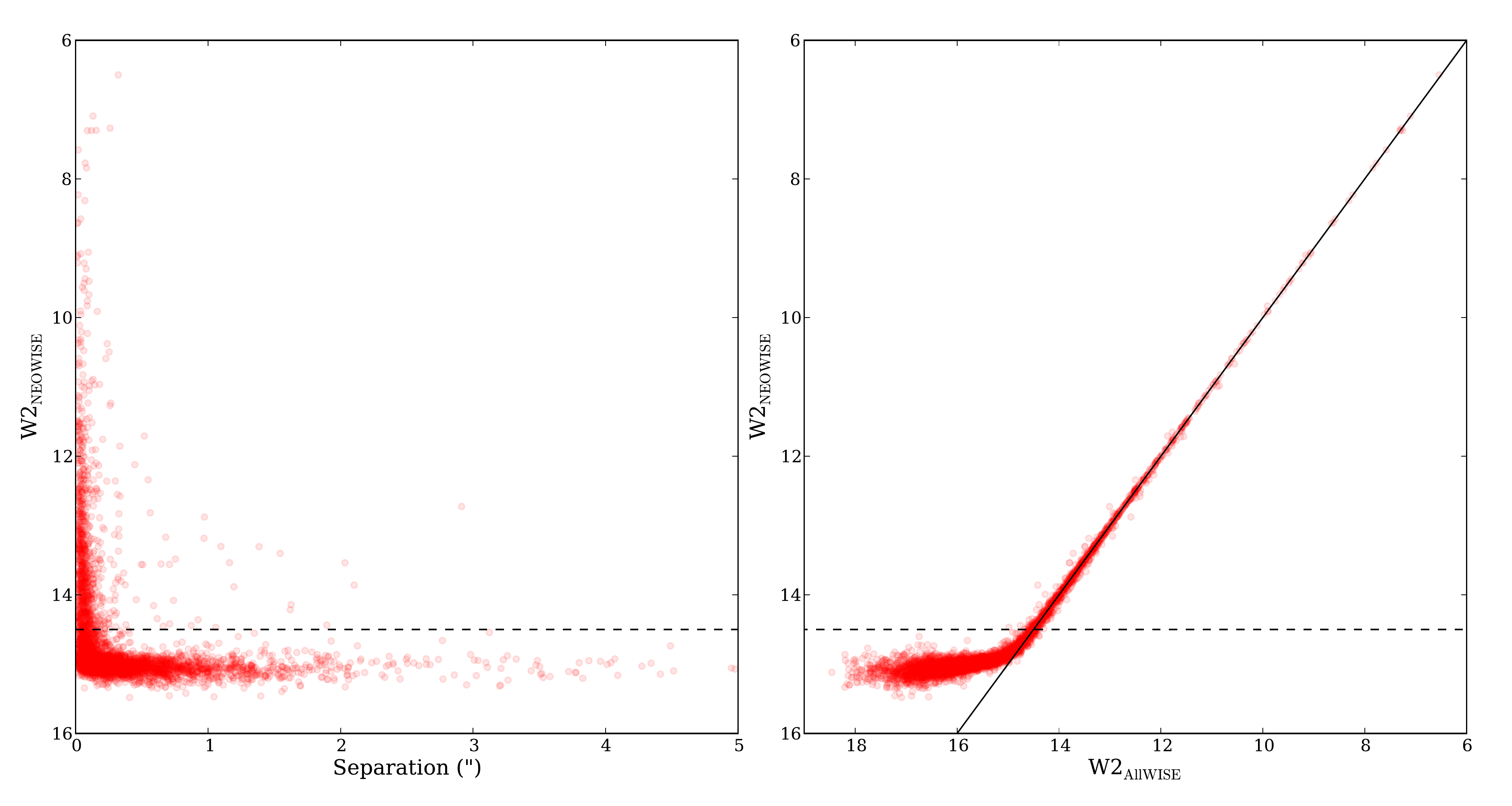}
\caption{{\it Left:} The NEOWISE W2 magnitude as a function of separation between the AllWISE and NEOWISE source catalog positions for a random sample of 5,000 objects.  The dashed line indicates a NEOWISE W2 magnitude of 14.5.  {\it Right:} A comparison of the NEOWISE W2 and AllWISE W2 magnitudes.  The solid line indicates a ratio of unity.  The dashed line indicates a NEOWISE W2 magnitude of 14.5.}  
\end{figure*}

One of the principal goals of our NEOWISE proper motion survey is to search for cold, nearby brown dwarfs.  Since such objects are typically too faint to be detected in W1, we conducted our search using W2 data.  For the additional science goals (i.e., identifying nearby M and L type subdwarfs), the difference between W1 and W2 is small (W1$-$W2 values for late type subdwarfs in Table 6 of \cite{kirk14} range from 0.10 to 0.56 mag), so a search in W2 alone will be sufficient to identify most objects of interest.  Our NEOWISE source table is assembled using the individual detections of each source with the aid of the STILTS tool set \citep{tay06}, a method very similar to that used in \cite{luh14a}.  STILTS is a set of command line tools designed specifically to handle large tables.  Sources in our NEOWISE source catalog are required to have at least five single detections within a 1\farcs5 radius, where the individual W2 magnitudes of each detection are off by no more than one magnitude from the median of all the other individual detections.  We consider all detections that occur within 10 days to form a single epoch.  A length of 10 days was chosen to account for the {\it WISE} telescope's Moon avoidance maneuvers.  We also require the individual detections to not be flagged as artifacts (i.e., {\it cc\_flags} $\neq$ [`D', `H', `O', `P']).  Lastly, we avoid the ecliptic poles (abs(elat) $\leq$ 85.0$\degr$) because the depth of coverage at the poles creates an extremely large amount of data for a relatively small area of the sky.  The final product of this process is a NEOWISE source catalog consisting of average right ascension, declination, W1, W2, and modified Julian Date values for each source.  

The accuracy of the astrometric and photometric measurements of sources in our NEOWISE source catalog decreases as objects become fainter in W2.  To identify the practical limits of our W2 magnitude search, we cross-matched a random sample of 5,000 entries from our NEOWISE source catalog (which should largely be unmoving background sources) with the AllWISE source catalog using a 5$\arcsec$ search radius.  The separation between the AllWISE and NEOWISE source positions as a function of the NEOWISE W2 magnitude, as well as a comparison between the NEOWISE and AllWISE W2 magnitudes, is shown in Figure 1.  The figure shows that below a NEOWISE W2 magnitude of $\sim$14.5, the positional and photometric NEOWISE values become unreliable.  We therefore make a W2 $\leq$ 14.5 magnitude cut to our final source catalog for our initial input sample.  

A typical proper motion survey will attempt to identify multiple detections of single objects at different epochs.  Our search strategy differs in that we identify high proper motion candidates as those that do not have a match at the previous epoch within a small search radius.  We identify potential high proper motion objects by cross-matching the positions of sources within our NEOWISE source catalog with the AllWISE catalog using a 1$\arcsec$ search radius, where those sources without matches are retained as potential high proper motion candidates.  In addition, each source is cross-matched with the AllWISE reject catalog and the 2MASS point source catalog \citep{cut03} using a 1$\arcsec$ search radius, again retaining only those without a match.   Cross-matching with the AllWISE reject catalog was necessary because we found that there are some instances where real objects near extremely bright sources can be flagged as artifacts, ending up in the AllWISE reject catalog instead of the AllWISE source catalog. In addition, we also found instances where there are real sources, usually blended with a slightly brighter source in the {\it WISE} images, that are in neither the AllWISE source or reject catalogs.  These sources are typically resolved in 2MASS and listed in the 2MASS point source catalog, hence the 2MASS 1$\arcsec$ search.  

Considering the $\sim$4 year time baseline between the first sky pass of NEOWISE and the first {\it WISE} epochs, our $1\arcsec$ search radius gives us a nominal minimum proper motion limit of $\sim$250 mas yr$^{-1}$.  We note that this limit is self-imposed, and that proper motions below this limit should also be detectable with the NEOWISE/AllWISE time baseline.  By not requiring a significance of motion threshold as in \cite{luh14a}, this survey probes to the faintest magnitude limits of what is possible with {\it WISE} single detections.  Using this method, the only upper boundary for detecting proper motions is the size of the {\it WISE} images in our finder charts.  Because the images are 2\arcmin $\times$ 2\arcmin, any object moving faster than $\sim$15$\arcsec$ yr$^{-1}$ (1\arcmin/$\sim$4 yr) would be beyond the boundary of the image.  Note that the two highest proper motion objects known (Barnard's Star -- 10\farcs4 yr$^{-1}$ \citep{bar16} and WISE J085510.83$-$071442.5 -- 8\farcs1 yr$^{-1}$ \citep{luh14d}) are both below this threshold and were recovered in our survey.      

When a NEOWISE source was found to not have a counterpart within 1$\arcsec$ in the AllWISE source catalog, the AllWISE reject catalog, or the 2MASS point source catalog, we created a finder chart by gathering available optical (DSS and SDSS), near-infrared (2MASS), and mid-infrared ({\it WISE} All-Sky) images.  Each individual finder chart was examined by-eye in an attempt to confirm each candidate's high proper motion by inspecting images at previous epochs.  A typical finder chart for a new high proper motion discovery is shown in Figure 2.  Sources that were discarded as spurious were typically blended or extended in nature.  Figure 3 shows an example of a high proper motion candidate that was determined to be a blended source (and therefore spurious) during the visual inspection process.  Over one million proper motion candidates were scrutinized in this way.  

\begin{figure*}
\plotone{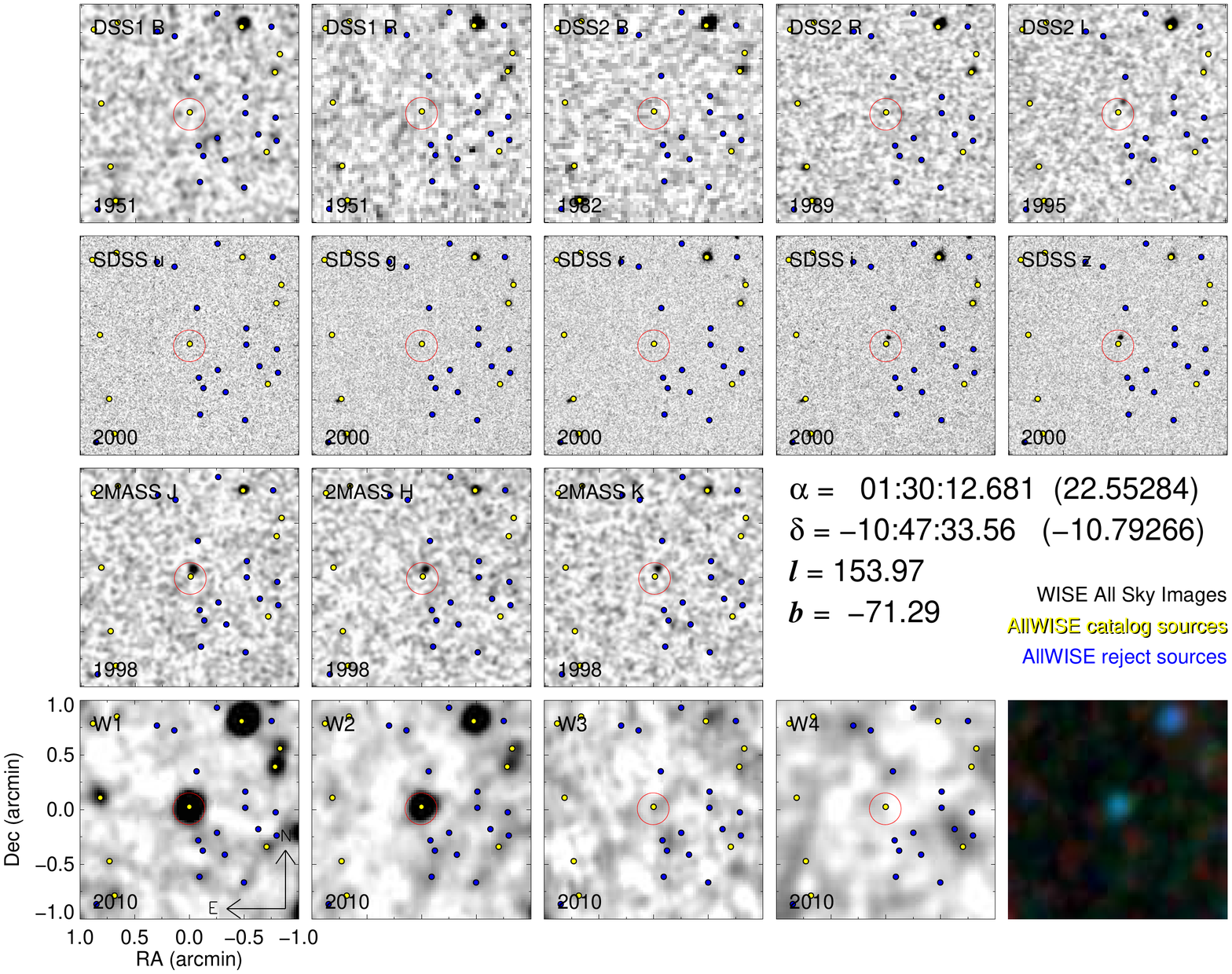}
\caption{Optical (DSS1, DSS2, SDSS), near-infrared (2MASS), and mid-infrared ({\it WISE} All-Sky) images of the newly discovered high proper motion object WISEA J013012.66$-$104732.4.  The red circle indicates the NEOWISE position of WISEA J013012.66$-$104732.4.  Yellow points indicate the positions of sources in the AllWISE source catalog, while blue points indicate the positions of sources in the AllWISE reject catalog. }  
\end{figure*}

\begin{figure}
\plotone{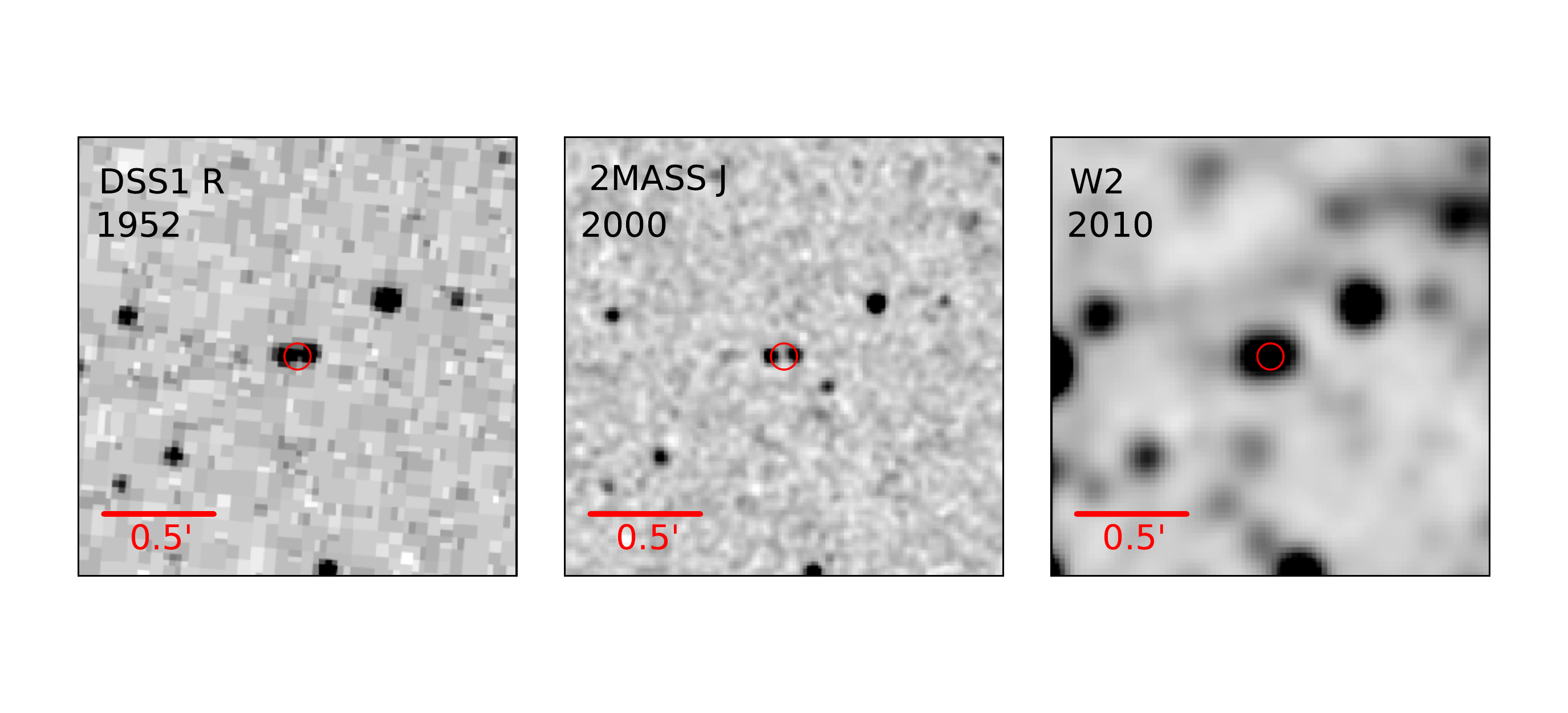}
\caption{2MASS and AllWISE images of a high proper motion candidate determined to be spurious.  The red circle denotes the NEOWISE position of the candidate. }  
\end{figure}

\section{Discussion}
\subsection{Survey Results}

A total of 20,548 high proper motion objects were found with the NEOWISE survey.  In order to determine if a confirmed proper motion source is known or is a new discovery, we rely primarily on the SIMBAD database.  We also checked catalogs of targeted searches for high proper motion objects (e.g., \citealt{pok04}, \citealt{lep05}, \citealt{dea07}, \citealt{boyd11}, \citealt{luh14a}, and \citealt{kirk14}).  Note that we only cross-match with catalogs made up of bona-fide proper motion sources, not unvetted lists of candidates (e.g., \citealt{gag15}). The vast majority of these objects were previously known to have significant proper motions.  The number of new high proper motion discoveries from this search totaled 1,006.     

Figure 4 shows the locations of all high proper motion objects identified in our NEOWISE survey.  The two gaps in coverage are due to a command timing anomaly that temporarily put the NEOWISE spacecraft in safe-mode (see the NEOWISE Data Release Explanatory Supplement for more details\footnote{http://wise2.ipac.caltech.edu/docs/release/neowise/expsup/sec1\_2.html}).  Similarly to \cite{kirk14}, most of the newly discovered high proper motion objects from this survey are located in the southern hemisphere, particularly near the Galactic center.  This is because, historically, there have been more targeted high proper motion searches in the northern hemisphere and the Galactic center is an exceptionally confused area because of its high density of stars.  We provide 2MASS and AllWISE associations, and 2MASS to AllWISE calculated proper motions for every newly discovered object in Table 1.  Proper motion uncertainties come from the 2MASS and AllWISE positional uncertainties. The same information for every previously known high proper motion object is provided in Table 2.  Upper limits for all magnitudes in all tables are at the 95\% confidence level\footnote{http://wise2.ipac.caltech.edu/docs/release/allwise/expsup/sec2\_1.html and http://www.ipac.caltech.edu/2mass/releases/allsky/doc/sec4\_4d.html)}.      

\begin{figure*}
\plotone{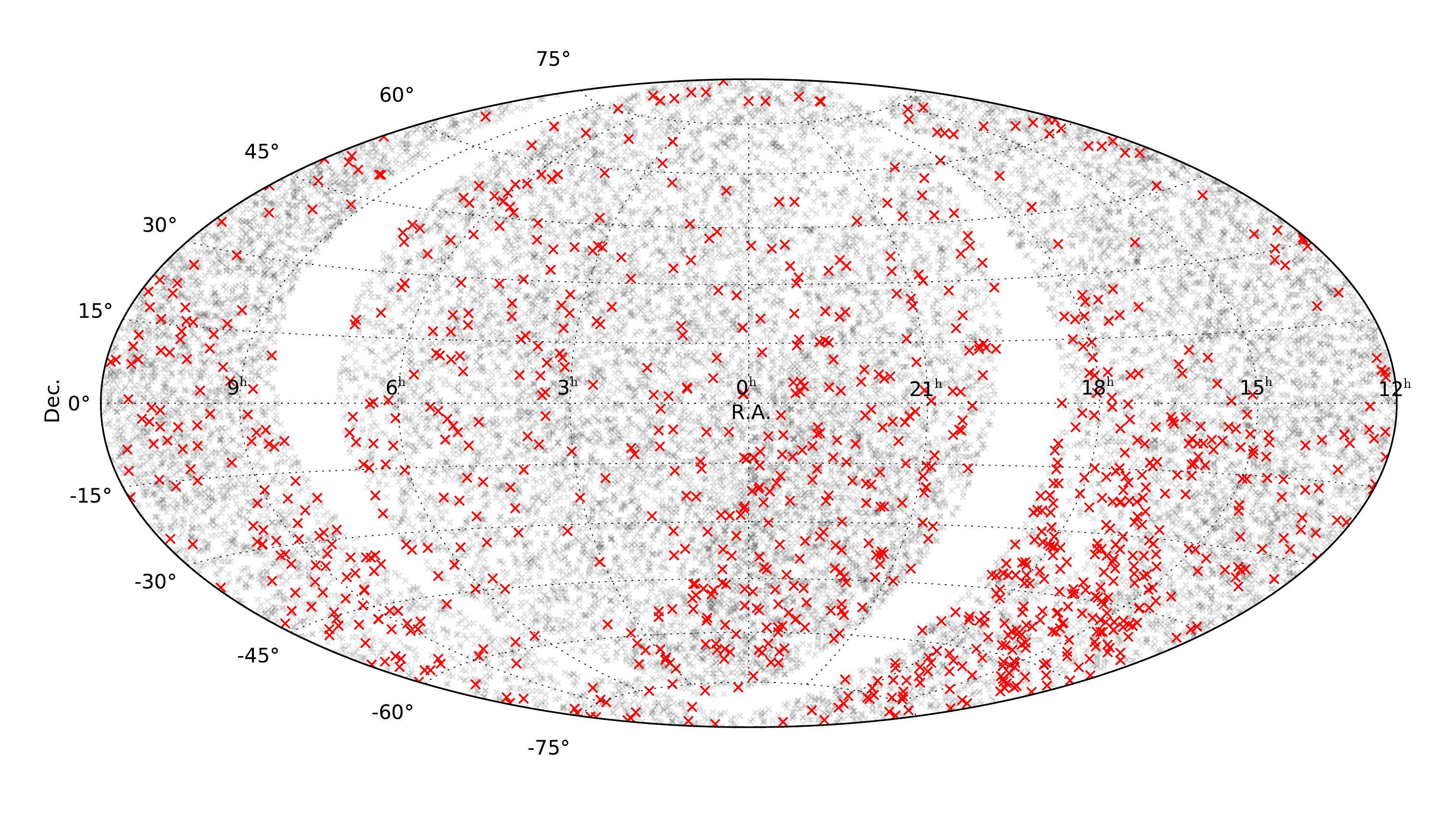}
\caption{The equatorial positions of every high proper motion source found with our NEOWISE proper motion survey.  Previously known objects with high proper motions are denoted by grey crosses, while new discoveries are plotted in red. The gaps in coverage are due to the NEOWISE command timing anomaly (see Sect.\ 3.1).}  
\end{figure*}

\begin{turnpage}
\begin{deluxetable*}{cccccccccc}
\tablecaption{New High Proper Motion Objects}
\tablehead{
\colhead{AllWISE } & W1 & W2 & 2MASS  & 2MASS J  &  2MASS H  & 2MASS K$_{\rm S}$ & $\mu$$_{\alpha}$ & $\mu$$_{\delta}$ & Type\tablenotemark{e} \\
Designation & (mag) & (mag) & Designation & (mag) & (mag) & (mag) & (mas yr$^{-1}$) & (mas yr$^{-1}$) & (photometric)}
\startdata
J000021.36$+$822940.1 & 14.057 $\pm$ 0.026 & 13.737 $\pm$ 0.031 & 00002008$+$8229395 & 15.060 $\pm$ 0.043 & 14.617 $\pm$ 0.062 & 14.316 $\pm$ 0.077 & 253.0 $\pm$ 13.2 & 68.1 $\pm$ 8.0 & 6.2\\
J000211.43$-$135749.0 & 12.890 $\pm$ 0.025 & 12.676 $\pm$ 0.025 & 00021121$-$1357480 & 13.973 $\pm$ 0.027 & 13.492 $\pm$ 0.035 & 13.138 $\pm$ 0.033 & 265.9 $\pm$ 7.1 & $-$79.0 $\pm$ 6.3 & 6.3\\
J000430.66$-$260402.3 & 15.211 $\pm$ 0.038 & 14.127 $\pm$ 0.044 & 00043065$-$2603596 & 16.487 $\pm$ 0.133 & 15.587 $\pm$ 0.129 & $>$15.523 & 11.9 $\pm$ 15.3 & $-$229.6 $\pm$ 13.8 & 20.5\\
J000458.47$-$133655.1\tablenotemark{a} & 15.120 $\pm$ 0.037 & 14.457 $\pm$ 0.056 & \dots & 16.841 $\pm$ 0.171 & 16.120 $\pm$ 0.207 & $>$15.410 & 431.3 $\pm$ 21.8 & $-$37.4 $\pm$ 20.3 & 16.9\\
J000502.05$+$021714.2 & 12.450 $\pm$ 0.023 & 12.228 $\pm$ 0.023 & 00050184$+$0217120 & 13.686 $\pm$ 0.029 & 13.056 $\pm$ 0.035 & 12.675 $\pm$ 0.023 & 328.1 $\pm$ 7.4 & 216.6 $\pm$ 7.3 & 6.8\\
J000534.07$-$475033.0 & 14.694 $\pm$ 0.030 & 14.412 $\pm$ 0.046 & 00053380$-$4750280 & 15.816 $\pm$ 0.077 & 15.129 $\pm$ 0.094 & 14.858 $\pm$ 0.118 & 248.3 $\pm$ 11.9 & $-$460.8 $\pm$ 10.3 & 5.0\\
J000536.63$-$263311.8 & 14.924 $\pm$ 0.033 & 14.261 $\pm$ 0.047 & 00053630$-$2633123 & 17.171 $\pm$ 0.225 & 15.849 $\pm$ 0.165 & 15.191 $\pm$ 0.154 & 384.0 $\pm$ 22.8 & 39.8 $\pm$ 20.5 & 17.1\\
J000551.34$+$021616.0 & 13.580 $\pm$ 0.026 & 13.351 $\pm$ 0.031 & 00055109$+$0216145 & 14.614 $\pm$ 0.024 & 14.009 $\pm$ 0.049 & 13.699 $\pm$ 0.059 & 382.7 $\pm$ 9.5 & 141.2 $\pm$ 8.6 & $<$5\\
J000603.34$-$522744.3 & 13.604 $\pm$ 0.026 & 13.340 $\pm$ 0.029 & 00060303$-$5227433 & 15.753 $\pm$ 0.070 & 14.794 $\pm$ 0.080 & 14.058 $\pm$ 0.061 & 282.4 $\pm$ 8.5 & $-$97.7 $\pm$ 8.4 & 13.5\\
J000627.85$+$185728.8 & 14.111 $\pm$ 0.027 & 13.785 $\pm$ 0.039 & 00062779$+$1857320 & 17.138 $\pm$ 0.206 & 15.686 $\pm$ 0.152 & 14.836 $\pm$ 0.084 & 72.2 $\pm$ 15.7 & $-$289.1 $\pm$ 14.8 & 16.1\\
J000856.39$-$281321.7 & 14.119 $\pm$ 0.027 & 13.636 $\pm$ 0.037 & 00085614$-$2813211 & 16.727 $\pm$ 0.137 & 15.664 $\pm$ 0.139 & 15.049 $\pm$ 0.131 & 284.3 $\pm$ 16.0 & $-$54.7 $\pm$ 13.6 & 18.0\\
J001320.75$+$271020.4 & 14.538 $\pm$ 0.030 & 14.356 $\pm$ 0.047 & 00132036$+$2710222 & 15.500 $\pm$ 0.060 & 14.884 $\pm$ 0.069 & 14.729 $\pm$ 0.112 & 404.8 $\pm$ 7.5 & $-$132.8 $\pm$ 7.5 & $<$5\\
J001422.56$-$333734.8 & 13.889 $\pm$ 0.026 & 13.612 $\pm$ 0.034 & 00142232$-$3337332 & 14.886 $\pm$ 0.045 & 14.377 $\pm$ 0.054 & 14.076 $\pm$ 0.068 & 254.7 $\pm$ 7.5 & $-$134.6 $\pm$ 7.5 & $<$5\\
J001643.97$+$230426.5 & 14.325 $\pm$ 0.029 & 13.667 $\pm$ 0.035 & 00164364$+$2304262 & $>$16.412 & 15.704 $\pm$ 0.138 & $>$14.973 & 386.1 $\pm$ 16.7 & 25.9 $\pm$ 15.8 & 18.9\\
J002000.06$-$070950.5 & 13.747 $\pm$ 0.027 & 13.435 $\pm$ 0.033 & 00195978$-$0709505 & 14.754 $\pm$ 0.034 & 14.200 $\pm$ 0.044 & 13.849 $\pm$ 0.044 & 358.5 $\pm$ 11.8 & $-$5.3 $\pm$ 11.8 & $<$5
\enddata
\tablenotetext{a}{2MASS photometry for this object is from the 2MASS Reject Catalog.}
\tablenotetext{b}{WISE photometry and positions for this object are from the AllWISE Reject Catalog.}
\tablenotetext{c}{WISE photometry and positions for this object are from the WISE All-Sky Catalog.}
\tablenotetext{d}{This object is a co-moving doubles that is resolved in 2MASS, but not in AllWISE.}
\tablenotetext{e}{Numerical spectral type estimates (e.g., 5 = M5, 15 = L5, 25 = T5)}
\tablecomments{(This table is available in its entirety in a machine-readable form in the online journal. A portion is shown here for guidance regarding its form and content.)}
\end{deluxetable*}
  
\begin{deluxetable*}{cccccccccc}
\tablecaption{Known High Proper Motion Objects}
\tablehead{
\colhead{AllWISE } & W1 & W2 & 2MASS  & 2MASS J  &  2MASS H  & 2MASS K$_{\rm S}$ & $\mu$$_{\alpha}$ & $\mu$$_{\delta}$ & \\
Designation & (mag) & (mag) & Designation & (mag) & (mag) & (mag) & (mas yr$^{-1}$) & (mas yr$^{-1}$)}
\startdata
J000012.91$-$545452.7 & 10.383 $\pm$ 0.023 & 10.379 $\pm$ 0.021 & 00001260$-$5454517 & 10.722 $\pm$ 0.020 & 10.475 $\pm$ 0.025 & 10.449 $\pm$ 0.023 & 249.3 $\pm$ 7.3 & $-$89.8 $\pm$ 6.5 \\
J000027.09$+$575404.9 & 11.734 $\pm$ 0.023 & 11.733 $\pm$ 0.023 & 00002657$+$5754025 & 12.497 $\pm$ 0.024 & 12.010 $\pm$ 0.031 & 11.855 $\pm$ 0.028 & 381.1 $\pm$ 6.7 & 219.1 $\pm$ 6.7 \\
J000028.04$-$412531.3 & 12.685 $\pm$ 0.023 & 12.548 $\pm$ 0.024 & 00002754$-$4125310 & 13.545 $\pm$ 0.026 & 12.974 $\pm$ 0.022 & 12.834 $\pm$ 0.032 & 506.6 $\pm$ 7.4 & $-$33.8 $\pm$ 6.6 \\
J000031.00$-$261352.0 & 9.328 $\pm$ 0.023 & 9.286 $\pm$ 0.020 & 00003078$-$2613533 & 10.400 $\pm$ 0.029 & 9.753 $\pm$ 0.031 & 9.523 $\pm$ 0.024 & 298.2 $\pm$ 7.9 & 124.8 $\pm$ 7.8 \\
J000031.98$+$650427.7 & 11.285 $\pm$ 0.023 & 11.144 $\pm$ 0.020 & 00003151$+$6504287 & 12.126 $\pm$ 0.022 & 11.558 $\pm$ 0.031 & 11.393 $\pm$ 0.021 & 274.3 $\pm$ 6.5 & $-$86.0 $\pm$ 6.5 \\
J000034.69$-$365006.8 & 10.809 $\pm$ 0.023 & 10.718 $\pm$ 0.020 & 00003429$-$3650079 & 11.698 $\pm$ 0.022 & 11.095 $\pm$ 0.023 & 10.912 $\pm$ 0.023 & 428.0 $\pm$ 7.3 & 105.7 $\pm$ 7.2 \\
J000037.66$+$420712.8 & 11.682 $\pm$ 0.024 & 11.614 $\pm$ 0.021 & 00003735$+$4207123 & 12.581 $\pm$ 0.022 & 11.958 $\pm$ 0.024 & 11.800 $\pm$ 0.024 & 300.2 $\pm$ 6.9 & 49.1 $\pm$ 6.1 \\
J000039.50$+$182921.9 & 7.506 $\pm$ 0.033 & 7.556 $\pm$ 0.020 & 00003925$+$1829198 & 8.443 $\pm$ 0.019 & 7.794 $\pm$ 0.023 & 7.639 $\pm$ 0.018 & 324.0 $\pm$ 8.8 & 193.9 $\pm$ 6.2 \\
J000040.37$+$162804.4 & 12.985 $\pm$ 0.024 & 12.738 $\pm$ 0.026 & 00004004$+$1628047 & 14.061 $\pm$ 0.031 & 13.519 $\pm$ 0.041 & 13.159 $\pm$ 0.037 & 441.2 $\pm$ 7.7 & $-$27.8 $\pm$ 6.8 \\
J000040.56$+$031339.3 & 12.849 $\pm$ 0.024 & 12.618 $\pm$ 0.026 & 00004044$+$0313424 & 13.711 $\pm$ 0.026 & 13.212 $\pm$ 0.031 & 12.964 $\pm$ 0.030 & 167.5 $\pm$ 12.7 & $-$307.3 $\pm$ 8.2 \\
J000044.53$-$502924.7 & 10.387 $\pm$ 0.023 & 10.230 $\pm$ 0.020 & 00004412$-$5029248 & 11.215 $\pm$ 0.030 & 10.726 $\pm$ 0.026 & 10.486 $\pm$ 0.024 & 394.2 $\pm$ 7.9 & 6.3 $\pm$ 7.0 \\
J000045.68$-$624345.6 & 8.992 $\pm$ 0.023 & 9.042 $\pm$ 0.020 & 00004539$-$6243437 & 9.885 $\pm$ 0.023 & 9.230 $\pm$ 0.023 & 9.070 $\pm$ 0.023 & 204.5 $\pm$ 6.9 & $-$186.0 $\pm$ 6.9 \\
J000047.16$-$351007.1 & 8.109 $\pm$ 0.022 & 8.072 $\pm$ 0.021 & 00004688$-$3510060 & 9.117 $\pm$ 0.029 & 8.480 $\pm$ 0.040 & 8.282 $\pm$ 0.027 & 343.2 $\pm$ 7.7 & $-$111.5 $\pm$ 6.8 \\
J000047.26$-$054116.7 & 12.797 $\pm$ 0.023 & 12.631 $\pm$ 0.027 & 00004707$-$0541187 & 13.789 $\pm$ 0.024 & 13.196 $\pm$ 0.022 & 12.927 $\pm$ 0.027 & 257.3 $\pm$ 12.5 & 184.0 $\pm$ 6.8 \\
J000052.23$+$143402.2 & 9.057 $\pm$ 0.024 & 9.009 $\pm$ 0.020 & 00005198$+$1434028 & 10.014 $\pm$ 0.019 & 9.382 $\pm$ 0.028 & 9.155 $\pm$ 0.023 & 345.4 $\pm$ 10.7 & $-$60.1 $\pm$ 7.2 
\enddata
\tablenotetext{a}{2MASS photometry for this object is from the 2MASS Reject Catalog.}
\tablenotetext{b}{WISE photometry and positions for this object are from the AllWISE Reject Catalog.}
\tablenotetext{c}{WISE photometry and positions for this object are from the WISE All-Sky Catalog.}
\tablenotetext{d}{This object is a co-moving doubles that is resolved in 2MASS, but not in AllWISE.}
\tablecomments{(This table is available in its entirety in a machine-readable form in the online journal. A portion is shown here for guidance regarding its form and content.)}
\end{deluxetable*}

\end{turnpage}

Seven objects were found to be moving upon visual inspection of their finder charts, but did not have a corresponding entry in any of the {\it WISE} catalogs based on coadded images (AllWISE, All-Sky, or Reject).  All of these objects but one (WISEA 19501894$+$2530402) are blended with a nearby, brighter source, which likely led to their omission from the {\it WISE} catalogs. Three of these objects are new discoveries, while the other four are known high proper motion objects.  2MASS designations and photometry for all objects without {\it WISE} detections are provided in Table 3.  

\begin{deluxetable*}{cccc}
\tablecaption{Objects Lacking an Entry in the WISE All-Sky and AllWISE Source Catalogs and Reject Tables}
\tablehead{
\colhead{2MASS } & 2MASS J  &  2MASS H  & 2MASS K$_{\rm S}$  \\
Designation & (mag) & (mag) & (mag) }
\startdata
\cutinhead{Known High Proper Motion Objects}
01570561$-$5925475 & 11.695$\pm$ 0.027 & 11.076 $\pm$ 0.027 & 10.777 $\pm$ 0.025\\
19501894$+$2530402 & 14.246 $\pm$ 0.027 & 13.648 $\pm$ 0.031 & 13.468 $\pm$ 0.045\\
21225632$+$3656001 & 13.712 $\pm$ 0.031 & 13.304 $\pm$ 0.036 & 13.117 $\pm$ 0.030\\
23164596$-$4047396 & 13.944 $\pm$ 0.026 & 13.447 $\pm$ 0.022 & 13.213 $\pm$ 0.034\\
\cutinhead{New Discoveries}
16161420$-$6146542 & 14.597 $\pm$ 0.035 & 14.138 $\pm$ 0.045 & 13.855 $\pm$ 0.054\\
16350859$-$3832440 & 13.168 $\pm$ 0.026 & 12.642 $\pm$ 0.032 & 12.343 $\pm$ 0.033\\
16411478$-$3215156 & 14.034 $\pm$ 0.028 & 13.403 $\pm$ 0.033 & 13.276 $\pm$ 0.038
\enddata
\end{deluxetable*}

There were also a total of 51 confirmed high proper motion objects for which there was no 2MASS counterpart.  Two of the objects are new discoveries, while the remainder are known T and Y dwarfs.  These two new discoveries are further discussed in Section 3.4.  AllWISE designations and photometry for all high proper motion objects without 2MASS detections are provided in Table 4.         

\begin{deluxetable}{ccc}
\tablecaption{Objects Lacking 2MASS Counterparts}
\tablehead{
\colhead{AllWISE } & W1 &  W2 \\
Designation & (mag) & (mag)  }
\startdata
\cutinhead{Known High Proper Motion Objects}
J000517.49$+$373720.4 & 16.764 $\pm$ 0.089 & 13.291 $\pm$ 0.031\\
J001505.88$-$461517.8 & 16.960 $\pm$ 0.101 & 14.218 $\pm$ 0.043\\
J005911.10$-$011401.1 & 16.899 $\pm$ 0.118 & 13.732 $\pm$ 0.039\\
J033605.04$-$014351.0 & 18.449 $\pm$ 0.470 & 14.557 $\pm$ 0.057\\
J045853.91$+$643452.6 & 16.439 $\pm$ 0.074 & 13.022 $\pm$ 0.027\\
J061213.88$-$303612.1 & 16.402 $\pm$ 0.061 & 14.038 $\pm$ 0.038\\
J062309.92$-$045624.5 & 16.845 $\pm$ 0.094 & 13.814 $\pm$ 0.035\\
J074457.24$+$562820.9 & 17.181 $\pm$ 0.118 & 14.531 $\pm$ 0.049\\
J075946.98$-$490454.0 & 16.997 $\pm$ 0.091 & 13.812 $\pm$ 0.032\\
J085510.74$-$071442.5 & 16.231 $\pm$ 0.064 & 13.704 $\pm$ 0.033\\
J090116.20$-$030636.0 & 17.188 $\pm$ 0.129 & 14.557 $\pm$ 0.054\\
J092906.76$+$040957.6 & 16.543 $\pm$ 0.083 & 14.254 $\pm$ 0.048\\
J094306.00$+$360723.3 & 18.176 $\pm$ 0.297 & 14.413 $\pm$ 0.048\\
J095047.31$+$011733.1 & 17.635 $\pm$ 0.182 & 14.507 $\pm$ 0.051\\
J101243.44$+$102059.8 & 16.319 $\pm$ 0.073 & 14.180 $\pm$ 0.047\\
J102557.67$+$030755.8 & 17.487 $\pm$ 0.194 & 14.136 $\pm$ 0.052\\
J102940.51$+$093514.1 & 16.780 $\pm$ 0.117 & 14.376 $\pm$ 0.074\\
J105257.95$-$194250.1 & 16.585 $\pm$ 0.084 & 14.111 $\pm$ 0.044\\
J111239.25$-$385700.5 & 17.478 $\pm$ 0.169 & 14.404 $\pm$ 0.048\\
J115013.85$+$630241.3 & 16.958 $\pm$ 0.089 & 13.405 $\pm$ 0.028\\
J115239.94$+$113406.9 & 16.825 $\pm$ 0.106 & 14.649 $\pm$ 0.063\\
J120444.60$-$015034.7 & 16.573 $\pm$ 0.088 & 14.672 $\pm$ 0.060\\
J121710.27$-$031112.1 & 15.267 $\pm$ 0.039 & 13.205 $\pm$ 0.034\\
J121756.92$+$162640.3 & 16.549 $\pm$ 0.082 & 13.128 $\pm$ 0.030\\
J125715.91$+$400854.2 & 16.672 $\pm$ 0.079 & 14.431 $\pm$ 0.045\\
J131106.21$+$012253.9 & 17.579 $\pm$ 0.198 & 14.703 $\pm$ 0.060\\
J131833.96$-$175826.3 & 17.513 $\pm$ 0.160 & 14.666 $\pm$ 0.058\\
J132233.63$-$234017.0 & 16.733 $\pm$ 0.087 & 13.960 $\pm$ 0.040\\
J140518.32$+$553421.3 & 18.765 $\pm$ 0.396 & 14.097 $\pm$ 0.037\\
J145715.01$+$581510.1 & 16.661 $\pm$ 0.059 & 14.417 $\pm$ 0.037\\
J150115.92$-$400418.2 & 16.091 $\pm$ 0.060 & 14.233 $\pm$ 0.043\\
J150411.81$+$102715.4 & 16.215 $\pm$ 0.055 & 14.063 $\pm$ 0.039\\
J151906.63$+$700931.3 & 17.084 $\pm$ 0.069 & 14.138 $\pm$ 0.031\\
J154151.65$-$225024.9\tablenotemark{a} & 16.736 $\pm$ 0.165 & 14.246 $\pm$ 0.063\\
J161215.92$-$342028.5 & 17.415 $\pm$ 0.199 & 13.984 $\pm$ 0.045\\
J161441.47$+$173935.4 & 18.174 $\pm$ 0.266 & 14.226 $\pm$ 0.040\\
J165311.03$+$444422.7 & 16.485 $\pm$ 0.048 & 13.824 $\pm$ 0.029\\
J181210.83$+$272144.2 & 17.468 $\pm$ 0.143 & 14.196 $\pm$ 0.039\\
J182831.08$+$265037.6 & $>$18.248   & 14.353 $\pm$ 0.045\\
J184124.74$+$700038.2 & 16.436 $\pm$ 0.044 & 14.355 $\pm$ 0.033\\
J201404.11$+$042409.0 & 17.296 $\pm$ 0.168 & 14.956 $\pm$ 0.069\\
J201920.75$-$114807.5 & 17.256 $\pm$ 0.152 & 14.305 $\pm$ 0.052\\
J205628.88$+$145953.6 & 16.480 $\pm$ 0.075 & 13.839 $\pm$ 0.037\\
J210200.14$-$442919.9 & 16.951 $\pm$ 0.111 & 14.139 $\pm$ 0.043\\
J215918.90$+$030502.4 & 14.887 $\pm$ 0.034 & 14.278 $\pm$ 0.048\\
J220905.75$+$271143.6 & $>$18.831  & 14.770 $\pm$ 0.055\\
J225540.75$-$311842.0 & 16.550 $\pm$ 0.079 & 14.161 $\pm$ 0.045\\
J232035.37$+$144830.1 & 16.588 $\pm$ 0.082 & 14.341 $\pm$ 0.057\\
J232519.55$-$410535.1 & 17.064 $\pm$ 0.114 & 14.108 $\pm$ 0.040\\ 
\cutinhead{New Discoveries}
J030919.70$-$501614.2 & 16.465 $\pm$ 0.057 & 13.631 $\pm$ 0.031\\
J133300.03$-$160754.4 & 17.698 $\pm$ 0.194 & 14.943 $\pm$ 0.069
\enddata
\tablenotetext{a}{WISE photometry and positions for WISE J154151.65$-$225024.9 are from the WISE All-Sky Catalog.} 
\end{deluxetable}

\subsection{Categorizing Discoveries}

In order to identify the most interesting objects for follow-up spectroscopic observations (e.g., nearby objects and late-type subdwarfs), we attempted to estimate the approximate spectral type of each new discovery using the available 2MASS and AllWISE photometry.  This was accomplished by using the k-Nearest Neighbors classification scheme described in Appendix A.  The last column of Table 1 gives the estimated numerical type for each new discovery (e.g., 5 = M5, 15 = L5, 25 = T5).  We chose to list numerical types to ensure that these estimates are not mistaken for actual spectral types determined from optical or near-infrared spectroscopy.  Note that the earliest estimated types from our classification scheme are M0, so any object with an earlier spectral type than M0 will likely be classified as early M using this method.  However, as we are most interested in late-type dwarfs (spectral types L and T), this does not affect our follow-up target prioritization.  For this reason, we only provide final photometric types for objects with estimated types later than M5.  Uncertainties for these types are typically $\sim$2 subtypes (see Appendix A).  

Figure 5 shows the $J-K_S$ vs.\ $J-W2$ color-color diagram for all high proper motion objects found during this survey.  The large cluster of sources at $J-K_S$ $\sim$0.7 and $J-W2$ $\sim$1.3 are early to mid-M dwarfs, which make up the vast majority of our new discoveries.  Several new discoveries at the edges of the main M dwarf clump are classified as having types later than L0, which may show that our technique of estimating spectral types photometrically may have difficulty properly classifying color outliers (see Section 3.3).    

\begin{figure*}
\plotone{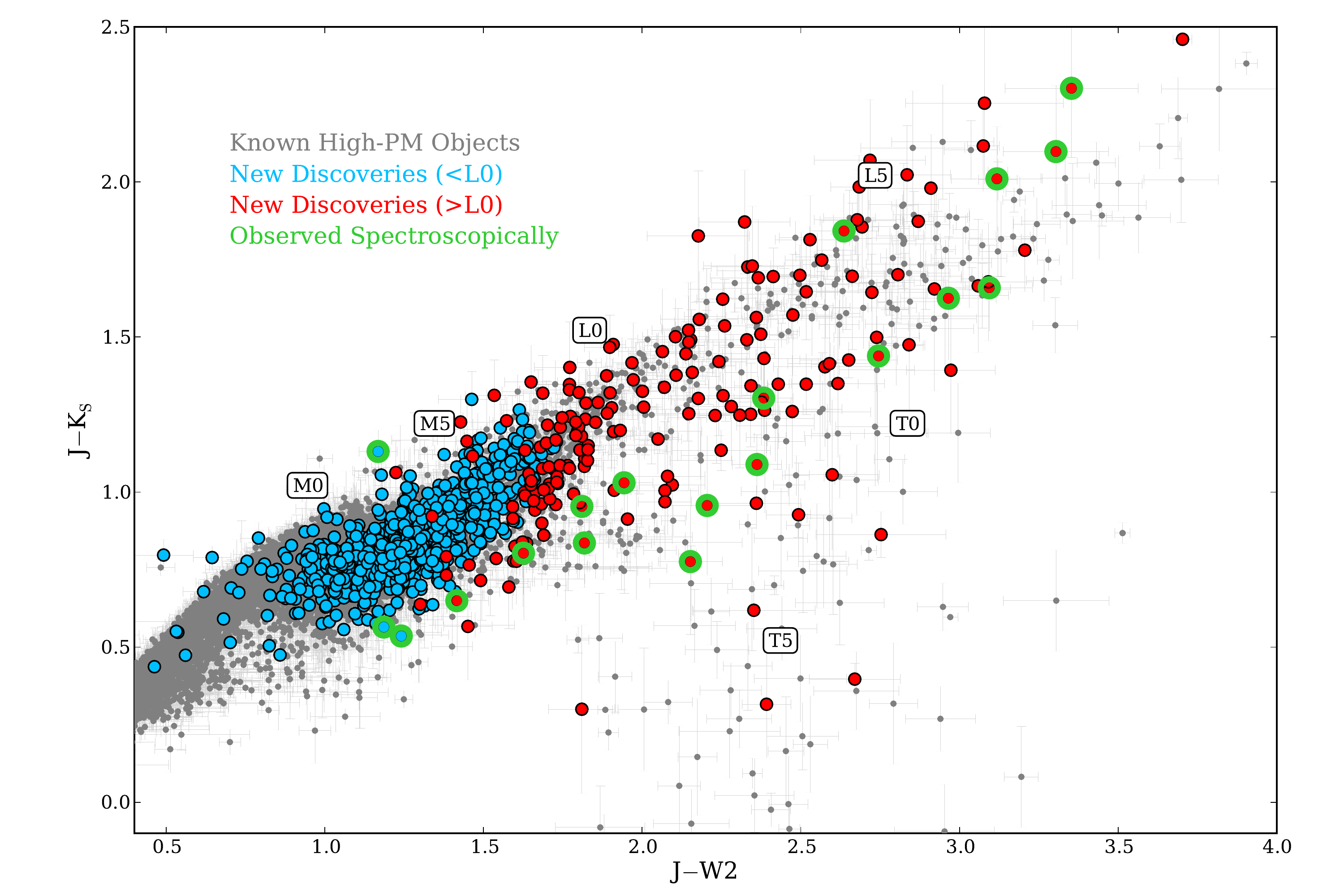}
\caption{$J-K_S$ vs. $J-W2$ color-color diagram for all high proper motion objects found with the NEOWISE survey.  Gray symbols are known objects from Table 2.  Light blue circles are new discoveries with spectral type estimates earlier than L0.  All other new discoveries are in red.  Objects that have been followed up spectroscopically are highlighted in green.  The approximate location of M0, M5, L0, L5, T0, and T5 spectral types in this color space is labeled for reference. Objects without 2MASS counterparts are not included in this figure.}  
\end{figure*}

\subsection{Common Proper Motion Pairs} 
During the vetting process of confirming high proper motion candidates, several new objects were noted to possibly be co-moving with a known high proper motion star.  Note that we did not perform a specific search for common proper motion pairs, only noted those that were noticed during the proper motion verification process of this survey.  Therefore, this list of common proper motion pairs is likely not exhaustive.  All potential new pairs are listed in Table 5, along with their 2MASS to AllWISE proper motions.  

\begin{deluxetable*}{ccccccc}
\tablecaption{New Common Proper Motion Pairs}
\tablehead{
\colhead{AllWISE } & $\mu$$_{\alpha}$ & $\mu$$_{\delta}$ & Known High-pm & $\mu$$_{\alpha}$ & $\mu$$_{\delta}$ & Separation\\
Designation & (mas yr$^{-1}$) & (mas yr$^{-1}$) & Star & (mas yr$^{-1}$) & (mas yr$^{-1}$) & (arcsec) }
\startdata
J003537.62$-$763750.7 & 201.6 $\pm$ 7.5 & $-$37.9 $\pm$ 6.1 & L 26$-$46 & 222.9 $\pm$ 7.2 & $-$22.4 $\pm$ 5.8 & 19.1\\
J014242.26$+$084824.3 & 129.0 $\pm$ 10.3 & $-$126.7 $\pm$ 8.6 & NLTT 5699 & 138.7 $\pm$ 9.2 & $-$161.2 $\pm$ 7.3 & 17.8\\
J040854.34$-$675105.0 & 216.9 $\pm$ 8.9 & 118.8 $\pm$ 7.4 & 2MASS J04083969$-$6750597 & 247.6 $\pm$ 7.5 & 189.7 $\pm$ 6.1 & 80.3\\
J050816.76$-$333021.9 & $-$65.8 $\pm$ 7.5 & $-$290.8 $\pm$ 6.7 & LTT 2180 & $-$61.3 $\pm$ 6.9 & $-$288.4 $\pm$6.1 & 13.9\\
J063228.30$+$264347.3 & 218.7 $\pm$ 7.5 & 35.9 $\pm$ 7.7 & G 103$-$38 & 250.4 $\pm$ 5.7 & 34.7 $\pm$ 5.5 & 25.6\\
J155017.09$-$862927.3\tablenotemark{a} & $-$296.1 $\pm$ 19.8 & $-$245.3 $\pm$ 10.2 & LHS 5302 & $-$334.6 $\pm$ 17.8 & $-$267.6 $\pm$ 7.4 & 12.2\\
J155039.10$-$504255.2 & 243.8 $\pm$ 6.5 & $-$73.3 $\pm$ 6.4 & GJ 599.1 & 271.5 $\pm$ 6.2 & $-$70.4 $\pm$ 6.1 & 276.4\\
J165906.03$-$784505.3 & $-$158.2 $\pm$ 8.1 & $-$186.7 $\pm$ 7.2 & NLTT 43745 & $-$188.1 $\pm$ 7.7 & $-$287.5 $\pm$ 6.8 & 20.9\\
J170027.83$-$220737.8\tablenotemark{a} & $-$68.1 $\pm$ 24.7 & $-$514.5 $\pm$ 55.7 & 2MASS J17002798$-$2207454 & $-$85.2 $\pm$ 6.4 & $-$466.5 $\pm$ 6.3 & 13.0\\
J184259.14$-$110921.6 & $-$208.4 $\pm$ 6.6 & $-$271.2 $\pm$ 6.4 & GJ 2139\tablenotemark{b} & $-$246 & $-$255 & 34.2\\  
J191648.99$+$470032.2 & $-$8.3 $\pm$ 7.2 & 282.3 $\pm$ 5.7 & HD 181096 & $-$19.1 $\pm$ 11.6 & 288.7 $\pm$ 10.2 & 40.6\\
J202422.29$-$063833.9 & 94.2 $\pm$ 8.7 & $-$205.0 $\pm$ 7.9 & 2MASS J20242285$-$0638224 & 75.9 $\pm$ 8.1 & $-$231.5 $\pm$  7.1 & 12.9\\
J203126.63$-$333515.9\tablenotemark{c} & 112.9 $\pm$ 6.3 & $-$138.6 $\pm$ 6.3 & WISEA J203126.61$-$333504.2\tablenotemark{c} & 71.0 $\pm$ 6.1 & $-$146.5 $\pm$ 5.9 & 12.5\\
J232308.63$-$631405.8 & 415.3 $\pm$ 8.3 & 23.5 $\pm$ 8.2 & 2MASS J23230415$-$6314327 & 417.9 $\pm$ 6.7 & 17.9 $\pm$ 6.6 & 37.1
\enddata
\tablenotetext{a}{The WISE designation for WISE J155017.09$-$862927.3 and WISE J170027.83$-$220737.8 are from the WISE All-Sky catalog}
\tablenotetext{b}{For GJ 2139, we quote the proper motions reported in \cite{stauff10}, as this object is a white dwarf not detected in any WISE catalog.}
\tablenotetext{c}{Both members of this pair are new discoveries.}
\end{deluxetable*}

Following \cite{luh14c}, we evaluate each pair using the companionship criterion proposed by \cite{lep07b}, which all pairs pass.  Table 6 provides additional information from the literature for the known high proper motion component of each pair.  For those that have a parallax measurement, we also include the projected separation between the pair in AU.  In each case, the NEOWISE discovery is the fainter component of the pair in W1 and W2 magnitude, with three exceptions.  WISEA J184259.14$-$110921.6 is a companion to the white dwarf GJ 2139, which was not detected in the AllWISE catalog.  WISEA J232308.63$-$631405.8 is slightly brighter in W1 and W2 than its known companion 2MASS J23230415$-$6314327.  Lastly, WISEA J203126.63$-$333515.9 and its companion WISEA J203126.61$-$333504.2 are both new discoveries from this survey.

\begin{deluxetable*}{cccccccccc}
\tablecaption{New Common Proper Motion Pair Properties}
\tablehead{
\colhead{AllWISE } & Type\tablenotemark{a} & Known High-pm & Sp. Type & Ref.\tablenotemark{b} & $\pi$ & Ref.\tablenotemark{b} & Fe/H & Ref.\tablenotemark{b} & Separation\\
Designation & (photometric) & Star & & & (arcsec) & & & & (AU) }
\startdata
J003537.62$-$763750.7 &  7.2 & L 26$-$46 & \dots & \dots & 11.06 $\pm$ 1.57 & 4 & -0.15 & 7 & 1726\\
J014242.26$+$084824.3 & 7.1 & NLTT 5699 & \dots & \dots & \dots & \dots & \dots & \dots & \dots \\
J040854.34$-$675105.0 & $<$5 & 2MASS J04083969$-$6750597 & \dots & \dots & \dots & \dots & \dots & \dots & \dots \\
J050816.76$-$333021.9 & 6.8 & LTT 2180 & \dots & \dots & \dots & \dots & \dots & \dots & \dots \\
J063228.30$+$264347.3 & 10.7 & G 103$-$38 & K5 & 1 & \dots & \dots & \dots & \dots & \dots \\
J155017.09$-$862927.3 & $<$5 & LHS 5302 & \dots & \dots & \dots & \dots & \dots & \dots & \dots \\
J155039.10$-$504255.2\tablenotemark{c} & $<$5 & GJ 599.1 & \dots & \dots & 23.17 $\pm$ 1.84 & 5 & -1.00 & 7 & 11929 \\
J165906.03$-$784505.3 &  9.0 & NLTT 43745 & \dots & \dots & \dots & \dots & -1.32 & 8 & \dots \\
J170027.83$-$220737.8\tablenotemark{c} &  7.1 & 2MASS J17002798$-$2207454 & \dots & \dots & \dots & \dots & \dots & \dots & \dots \\
J184259.14$-$110921.6 & $<$5 & GJ 2139 & DA4.9 & 2 & 53.0 $\pm$ 6.0 & 6 & \dots & \dots & 645\\  
J191648.99$+$470032.2 & $<$5 & HD 181096 & F6IV: & 3 & 23.79 $\pm$ 0.32 & 5 & -0.278 & 9 & 1706\\
J202422.29$-$063833.9 & 7.0 & 2MASS J20242285$-$0638224 & \dots & \dots & \dots & \dots & \dots & \dots & \dots \\
J203126.63$-$333515.9 & 6.7 & WISEA J203126.61$-$333504.2 & \dots & \dots & \dots & \dots & \dots & \dots & \dots \\
J232308.63$-$631405.8 & 7.2 & 2MASS J23230415$-$6314327 & \dots & \dots & \dots & \dots & \dots & \dots & \dots 
\enddata
\tablenotetext{a}{Numerical spectral type estimates (e.g., 5 = M5, 15 = L5, 25 = T5)}
\tablenotetext{b}{References: (1) \cite{lee84}; (2) \cite{gia11}; (3) \cite{hoff91}; (4) \cite{kor13}; (5) \cite{van07}; (6) \cite{gli91}; (7) \cite{amm06}; (8) \cite{ryan91}; (9) \cite{tay05}  }
\tablenotetext{c}{The WISE designation for WISE J155017.09$-$862927.3 and WISE J170027.83$-$220737.8 are from the WISE All-Sky catalog}
\end{deluxetable*}

\subsection{Nearby Objects}

Using our estimated spectral types (with a $\pm$ 2 subtype uncertainty), W2 magnitudes, and the absolute magnitude$-$spectral type relation for W2 from \cite{dup12}, we estimate a distance range to every new discovery with an estimated spectral type later than M5 in an attempt to identify new nearby objects.  We use the W2 magnitude because each object in our new discovery list is detected with AllWISE (with the exception of the three new objects in Table 3) and has a W2 magnitude $\lesssim$ 14.5, which corresponds to a signal-to-noise ration (S/N) of $\sim$15. We initially identified $\sim$70 objects with distance estimates $\leq$ 25 pc.  However, upon visual inspection of the finder charts, several high proper motion objects were found to be unresolved blends in the {\it WISE} images, which likely affected their photometry and led to erroneous spectral types.  We visually inspected the finder charts for each of the $\sim$70 objects potentially within 25 pc, flagging those that were blended in the {\it WISE} images.  All of these blends contain a high proper motion source and an unrelated, stationary background source (i.e., none are co-moving doubles).  Such blends can cause objects to be misclassified using our classification scheme, and will cause them to appear overluminous, and hence closer, than they actually are.  We omit all such blends from our list of potentially nearby sources.  The remaining 46 objects are listed in Table 7.  One object (WISEA J105515.71$-$735611.3) is estimated to be within $\sim$10 pc, with a photometric type estimate of 7.4 ($\sim$M7).  Three of these objects (WISEA J001643.97$+$230426.5, WISEA J003338.45$+$282732.4, and WISEA J010202.11$+$035541.4) have been followed-up spectroscopically and are discussed further in Section 5.2. 

\begin{deluxetable}{ccc}
\tablecaption{New Potential Nearby Objects}
\tablehead{
\colhead{AllWISE } & Type\tablenotemark{a} & Dist. \\
Designation & (photometric) & (pc)}
\startdata
J000856.39$-$281321.7 & 18.0 & 24$-$34\\
J001643.97$+$230426.5 & 18.9 & 23$-$31\\
J003338.45$+$282732.4 & 17.4 & 24$-$33\\
J010202.11$+$035541.4 & 18.9 & 23$-$31\\
J022721.93$+$235654.3 & 19.4 & 22$-$31\\
J030119.39$-$231921.1 & 20.5 & 24$-$33\\
J032309.12$-$590751.0 & 26.2 & 16$-$26\\
J034858.75$-$562017.8 & 22.6 & 24$-$33\\
J041743.13$+$241506.3 & 23.7 & 13$-$19\\
J053424.45$+$165255.0 & 15.4 & 18$-$25\\
J054455.54$+$063940.3 & 10.3 & 24$-$37\\
J060202.67$+$724235.4 & 18.7 & 23$-$31\\
J061429.77$+$383337.5 & 10.3 & 18$-$27\\
J083625.91$-$325034.5 & 5.2 & 22$-$35\\
J084254.56$-$061023.7 & 22.7 & 20$-$29\\
J085039.11$-$022154.3 & 16.3 & 21$-$ 30\\
J092740.70$-$500606.8 & 8.4 & 17$-$26\\
J093654.63$-$334620.5 & 9.9 & 22$-$34\\
J101944.62$-$391151.6 & 24.0 & 19$-$28\\
J105515.71$-$735611.3 & 7.4 & 8$-$12\\
J105811.69$-$583112.4 & 15.4 & 22$-$30\\
J111551.33$-$673135.5 & 8.4 & 13$-$21\\
J114117.13$-$790940.4 & 6.5 & 22$-$36\\
J121559.16$-$635351.8 & 7.0 & 22$-$35\\
J124138.43$-$643646.0 & 8.4 & 14$-$22\\
J130015.16$-$602417.2 & 6.0 & 16$-$26\\
J145640.16$-$535155.1 & 8.0 & 13$-$20\\
J150358.26$-$483505.0 & 8.3 & 19$-$30\\
J151029.95$-$604059.1 & 7.9 & 17$-$27\\
J154119.34$-$445055.8 & 8.0 & 22$-$34\\
J154209.42$-$515947.9 & 6.4 & 23$-$37\\
J164052.33$-$430750.7 & 5.8 & 12$-$20\\
J165057.66$-$221616.8 & 5.4 & 22$-$35\\
J165842.54$+$510334.9 & 13.7 & 23$-$33\\
J170234.91$-$670504.8 & 7.3 & 24$-$38\\
J171059.52$-$180108.7 & 5.2 & 24$-$37\\
J171105.08$-$275531.7 & 7.3 & 21$-$34\\
J171156.91$-$495441.1 & 8.1 & 20$-$31\\
J173551.56$-$820900.3 & 24.3 & 14$-$21\\
J174249.38$-$241101.6 & 7.6 & 10$-$15\\
J175546.92$-$340432.0 & 6.5 & 20$-$31\\
J183654.10$-$135926.2 & 8.7 & 20$-$31\\
J191011.03$+$563429.3 & 11.6 & 16$-$23\\
J201252.78$+$124633.3 & 6.5 & 17$-$26\\
J215620.63$-$532636.6 & 7.9 & 21$-$32\\
J225907.03$-$542036.9 & 7.0 & 17$-$28\\
\cutinhead{WISE-only Sources}
J030919.70$-$501614.2 & T7$-$T9 & 9$-$13 \\
J133300.03$-$160754.4 & T7$-$T9 & 17$-$24
\enddata
\tablenotetext{a}{Numerical spectral type estimates (e.g., 5 = M5, 15 = L5, 25 = T5)}
\end{deluxetable}

In addition, there are two objects from our discovery list that have no 2MASS counterpart (see Section 3.1 and Table 4).  Both of these discoveries (WISEA J030919.70$-$501614.2 and WISEA J133300.03$-$160754.4) have very red W1-W2 colors (2.83 and 2.76, respectively), indicating spectral types $\geq$T7 for both objects \citep{kirk12}.  Distance estimates for these two new late T dwarfs are 9$-$13 and 17$-$24 pc, respectively, for WISEA J030919.70$-$501614.2 and WISEA J133300.03$-$160754.4 based on spectral type estimates of T7 to T9.  These objects are also listed in Table 7.
  
\subsection{Late Type Subdwarfs}

Subdwarfs are low metallicity objects that are typically associated with the halo population, often having significantly larger tangential velocity ($V_{\rm tan}$) values than the field population.  We select candidate late-type subdwarfs using two different strategies.  First, as noted in \cite{kirk14}, many early L-type subdwarfs occupy a distinct region of J$-$K$_{\rm S}$ vs.\ J$-$W2 color space blueward of the main clump of mostly M-type main sequence stars in J$-$K$_{\rm S}$ color.  Figure 6 shows a close-up view of this region along with the known early-type L subdwarfs from \cite{kirk14}.  We chose as subdwarf candidates those objects which lie blueward in J$-$K$_{\rm S}$ color from the main clump of discoveries from this survey.  Specifically, subdwarf candidates are those with a J$-$W2 color between 0.9 and 1.25 mag and a J$-$K$_{\rm S}$ color less than 0.6 mag, a J$-$W2 color between 1.25 and 1.65 mag and a J$-$K$_{\rm S}$ value less than 0.8$\times$(J$-$W2) $-$ 0.4, or a J$-$W2 color between 1.65 and 1.9 mag and a J$-$K$_{\rm S}$ color less than 0.92 mag, as shown in Figure 6.  Nine of the fourteen early-L subdwarfs from \cite{kirk14} in the figure meet the above criteria.  Thirty-one subdwarf candidates were selected based on their colors and are listed in Table 8.  While we include estimated spectral types for these objects in the table, we note that our spectral type estimation technique presented in Appendix A is predicated on the object in question having colors typical of a normal star or brown dwarf.  Therefore, objects with colors distinct from those of normal late-type stars and brown dwarfs (such as subdwarfs) will likely be mistyped.  

\begin{figure*}
\plotone{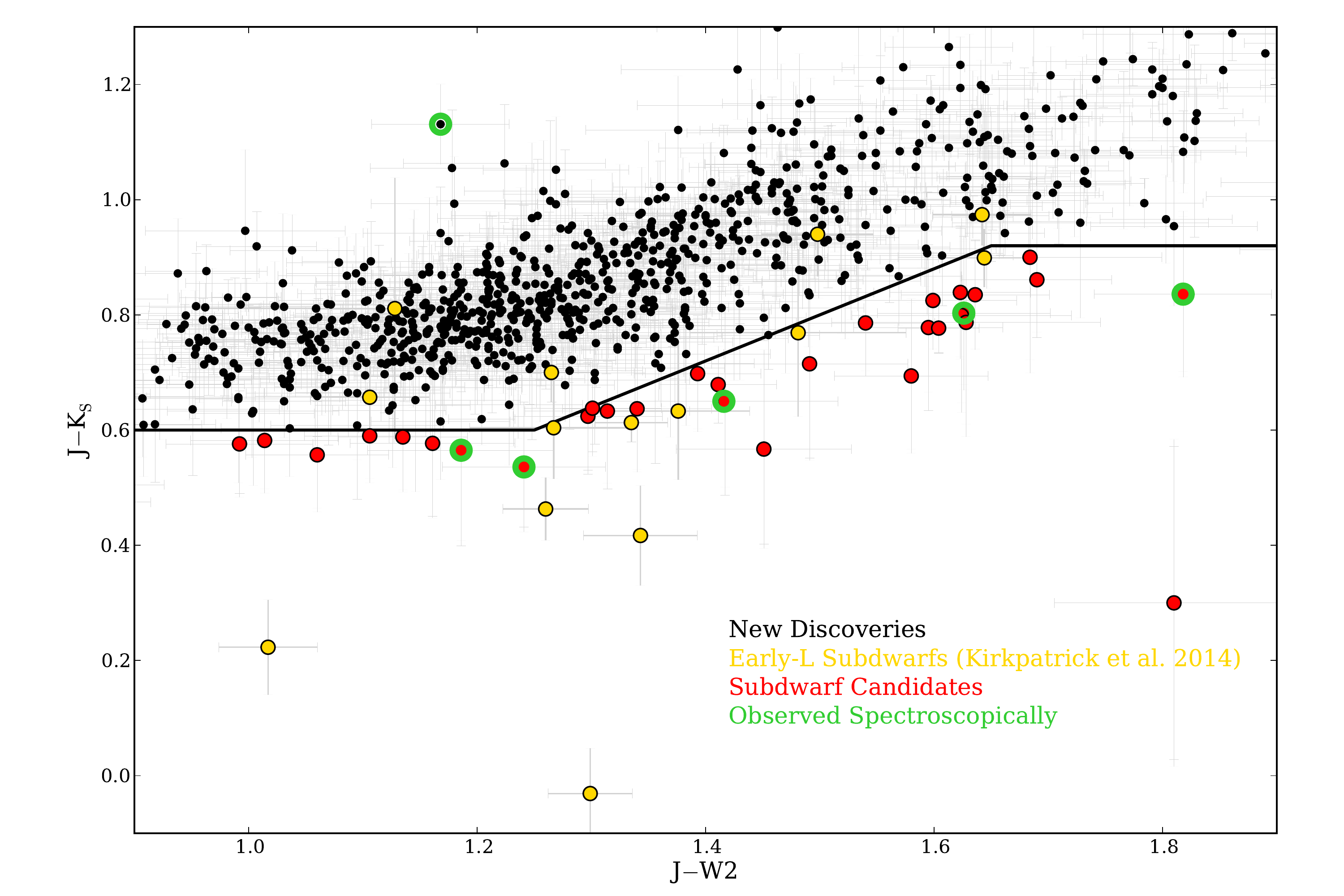}
\caption{J$-$K$_{\rm S}$ vs.\ J$-$W2 color-color diagram showing our color-selected subdwarf candidates.  Early-L subdwarfs from \cite{kirk14} are plotted in gold. The solid line denotes our subdwarf color-candidate selection criteria.} Objects that have been followed up spectroscopically are highlighted in green.    
\end{figure*}

\begin{deluxetable}{cccc}
\tablecaption{Color-Selected Subdwarf Candidates}
\tablehead{
\colhead{AllWISE } & J$-$K$_{\rm S}$ & J$-$W2 & Type\tablenotemark{a} \\
Designation & (mag) & (mag) & (photometric)}
\startdata
J003449.93$+$551352.8 & 0.59 $\pm$ 0.08 & 1.11 $\pm$ 0.06 & 7.0 \\
J011639.05$-$165420.5 & 0.84 $\pm$ 0.14 & 1.82 $\pm$ 0.08 & 15.8 \\
J013012.66$-$104732.4 & 0.80 $\pm$ 0.13 & 1.63 $\pm$ 0.08 & 10.9 \\
J025635.13$-$663443.9 & 0.84 $\pm$ 0.10 & 1.64 $\pm$ 0.06 & 10.2 \\
J050750.72$-$034245.8 & 0.78 $\pm$ 0.11 & 1.60 $\pm$ 0.07 & 12.0 \\
J052452.57$+$463202.9 & 0.58 $\pm$ 0.09 & 0.99 $\pm$ 0.06 & 8.2 \\
J063228.30$+$264347.3 & 0.82 $\pm$ 0.09 & 1.60 $\pm$ 0.07 & 10.7 \\
J094812.21$-$290329.5 & 0.69 $\pm$ 0.13 & 1.58 $\pm$ 0.07 & 11.8 \\
J094904.92$+$023251.4 & 0.90 $\pm$ 0.20 & 1.68 $\pm$ 0.12 & 15.5 \\
J101944.62$-$391151.6 & 0.30 $\pm$ 0.28 & 1.81 $\pm$ 0.10 & 24.0 \\
J105617.57$-$465101.8 & 0.63 $\pm$ 0.14 & 1.31 $\pm$ 0.07 & 8.0 \\
J112152.91$-$264937.3 & 0.59 $\pm$ 0.10 & 1.13 $\pm$ 0.06 & 6.4 \\
J114553.61$-$250657.1 & 0.56 $\pm$ 0.17 & 1.19 $\pm$ 0.08 & 8.4 \\
J120751.17$+$302808.9 & 0.79 $\pm$ 0.09 & 1.54 $\pm$ 0.06 & 10.6 \\
J122355.12$+$551050.3 & 0.57 $\pm$ 0.17 & 1.45 $\pm$ 0.08 & 12.2 \\
J124516.66$+$601607.5 & 0.58 $\pm$ 0.13 & 1.16 $\pm$ 0.07 & 9.4 \\
J143559.87$-$443930.9 & 0.64 $\pm$ 0.11 & 1.34 $\pm$ 0.07 & 9.9 \\
J143942.79$-$110045.4 & 0.80 $\pm$ 0.17 & 1.62 $\pm$ 0.10 & 11.8 \\
J144056.64$-$222517.8 & 0.62 $\pm$ 0.10 & 1.30 $\pm$ 0.07 & 9.1 \\
J155437.88$-$362534.4 & 0.68 $\pm$ 0.07 & 1.41 $\pm$ 0.05 & 9.8 \\
J162046.30$-$485952.1 & 0.78 $\pm$ 0.04 & 1.60 $\pm$ 0.04 & 12.5 \\
J163155.36$+$671549.3 & 0.70 $\pm$ 0.10 & 1.39 $\pm$ 0.06 & 9.2 \\
J174006.68$-$733720.4 & 0.71 $\pm$ 0.17 & 1.49 $\pm$ 0.09 & 11.3 \\
J180839.55$+$070021.7 & 0.79 $\pm$ 0.20 & 1.63 $\pm$ 0.12 & 14.7 \\
J182010.20$+$202125.8 & 0.58 $\pm$ 0.09 & 1.01 $\pm$ 0.06 & 7.4 \\
J213512.09$-$043155.0 & 0.84 $\pm$ 0.12 & 1.62 $\pm$ 0.06 & 10.4 \\
J221126.37$-$192207.4 & 0.65 $\pm$ 0.16 & 1.42 $\pm$ 0.10 & 10.2 \\
J221737.41$-$355242.7 & 0.64 $\pm$ 0.08 & 1.30 $\pm$ 0.06 & 10.3 \\
J232656.09$-$181504.5 & 0.54 $\pm$ 0.11 & 1.24 $\pm$ 0.07 & 9.5 \\
J234404.85$-$250042.2 & 0.86 $\pm$ 0.10 & 1.69 $\pm$ 0.07 & 11.8 \\
J234812.74$-$530649.7 & 0.56 $\pm$ 0.10 & 1.06 $\pm$ 0.06 & 8.3 
\enddata
\tablenotetext{a}{Numerical spectral type estimates (e.g., 5 = M5, 15 = L5, 25 = T5)}
\end{deluxetable}

Besides being distinguishable in color space, subdwarfs often show kinematics distinct from that of the field population.  These kinematic differences cause subdwarfs to stand out prominently in reduced proper motion diagrams, where the reduced proper motion is defined as H$_m$ = $m$ $+$ 5log($\mu$) $+$ 5, where $m$ is a particular photometric band and $\mu$ is the total proper motion.  Figure 7 shows a reduced proper motion diagram for all of the discoveries from this NEOWISE search, as well as the known late-type subdwarfs from \cite{kirk14}.  We select candidate subdwarfs as either having H$_{\rm J}$ values greater than 18.7 mag and J$-$W2 values less than 1.8 or having H$_{\rm J}$ values greater than $\frac{3.8}{1.5}$$\times$(J$-$W2) $+$ 14.14, as shown in the figure.  These two selection criteria pick out 18 of the 21 known, late-type subdwarfs from \cite{kirk14} that have available J-band photometry and proper motion measurements. Thirty-one objects were selected as subdwarf candidates based on the reduced proper motion diagram positions.  These subdwarf candidates are listed in Table 9.  Four objects (WISEA J011639.05$-$165420.5, WISEA J094812.21$-$290329.5, WISEA J094904.92$+$023251.4, and WISEA J101944.62$-$391151.6) are common to both the reduced proper motion and color-selected subdwarf lists.

\begin{deluxetable}{cccc}
\tablecaption{Reduced Proper Motion Subdwarf Candidates}
\tablehead{
\colhead{AllWISE } & H$_{\rm J}$ & J$-$W2 & Type\tablenotemark{a}  \\
Designation & (mag) & (mag) & (photometric) }
\startdata
J000534.07$-$475033.0 & 19.41 $\pm$ 0.10 & 1.40 $\pm$ 0.09 & 5.0\\
J004555.13$+$795848.7 & 20.94 $\pm$ 0.21 & 2.43 $\pm$ 0.04 & 16.8\\
J010134.83$+$033616.0 & 19.97 $\pm$ 0.17 & 1.48 $\pm$ 0.07 & 7.0\\
J011639.05$-$165420.5 & 19.61 $\pm$ 0.09 & 1.82 $\pm$ 0.08 & 15.8\\
J013042.06$-$064705.1 & 19.38 $\pm$ 0.10 & 1.36 $\pm$ 0.09 & 6.3\\
J022045.20$-$550622.7 & 19.31 $\pm$ 0.10 & 1.81 $\pm$ 0.10 & 10.4\\
J025612.30$+$684752.6 & 18.92 $\pm$ 0.08 & 1.40 $\pm$ 0.08 & 6.8\\
J030421.32$-$394550.8 & 19.28 $\pm$ 0.10 & 1.59 $\pm$ 0.07 & 13.7\\
J032309.12$-$590751.0 & 21.18 $\pm$ 0.31 & 2.35 $\pm$ 0.19 & 26.2\\
J033346.88$+$385152.6 & 19.10 $\pm$ 0.08 & 1.72 $\pm$ 0.09 & 11.6\\
J044111.37$+$285338.2 & 19.48 $\pm$ 0.08 & 1.14 $\pm$ 0.09 & $<$5.0\\
J055115.91$+$535607.9 & 20.10 $\pm$ 0.11 & 1.37 $\pm$ 0.09 & 5.5\\
J084903.52$-$511850.3 & 20.32 $\pm$ 0.17 & 2.18 $\pm$ 0.16 & 12.4\\
J092453.76$+$072306.0 & 19.05 $\pm$ 0.10 & 1.26 $\pm$ 0.10 & 5.9\\
J094812.21$-$290329.5 & 18.76 $\pm$ 0.07 & 1.58 $\pm$ 0.07 & 11.8\\
J094904.92$+$023251.4 & 19.44 $\pm$ 0.16 & 1.68 $\pm$ 0.12 & 15.5\\
J095230.79$-$282842.2 & 18.96 $\pm$ 0.07 & 1.29 $\pm$ 0.05 & 5.3\\
J101944.62$-$391151.6 & 19.62 $\pm$ 0.18 & 1.81 $\pm$ 0.10 & 24.0\\
J112158.76$+$004412.3 & 19.00 $\pm$ 0.09 & 1.66 $\pm$ 0.09 & 8.6\\
J121914.75$+$081027.0 & 19.03 $\pm$ 0.10 & 1.21 $\pm$ 0.10 & $<$5.0\\
J122042.20$+$620528.3 & 19.12 $\pm$ 0.08 & 1.28 $\pm$ 0.06 & 6.3\\
J122402.69$-$714057.3 & 19.07 $\pm$ 0.73 & 1.58 $\pm$ 0.04 & 8.5\\
J123513.87$-$045146.5 & 18.80 $\pm$ 0.08 & 1.15 $\pm$ 0.09 & 5.1\\
J132240.10$-$331836.4 & 18.75 $\pm$ 0.10 & 1.77 $\pm$ 0.11 & 11.0\\
J133520.09$-$070849.3 & 19.22 $\pm$ 0.11 & 1.77 $\pm$ 0.11 & 12.3\\
J152548.25$-$374651.2 & 18.78 $\pm$ 0.07 & 1.02 $\pm$ 0.07 & $<$5.0\\
J155225.22$+$095155.5 & 18.81 $\pm$ 0.10 & 1.37 $\pm$ 0.10 & 7.9\\
J160502.46$-$303205.9 & 18.74 $\pm$ 0.08 & 1.25 $\pm$ 0.09 & $<$5.0\\
J171643.78$+$200616.1 & 19.62 $\pm$ 0.08 & 1.23 $\pm$ 0.06 & $<$5.0\\
J171651.56$-$163912.5 & 19.30 $\pm$ 0.09 & 1.39 $\pm$ 0.06 & 5.9\\
J214338.47$-$170723.8 & 18.84 $\pm$ 0.10 & 1.64 $\pm$ 0.11 & 10.6
\enddata
\tablenotetext{a}{Numerical spectral type estimates (e.g., 5 = M5, 15 = L5, 25 = T5)}
\end{deluxetable}

We have obtained follow-up spectra for five of these candidates (WISEA J011639.05$-$165420.5, WISEA J013012.66$-$104732.4, WISEA J114553.61$-$250657.1, WISEA J221126.37$-$192207.4 and WISEA J232656.09$-$181504.5).  Each is discussed further in Section 5.2.  We also observed one object that stood out in J$-$K$_{\rm S}$ and J$-$W2 color space (WISEA J172602.92$-$034211.7; J$-$K$_{\rm S}$ =  1.13 mag, J$-$W2 = 1.17 mag, see Figure 6) that turned out to be an early M-type subdwarf.      

\begin{figure}
\plotone{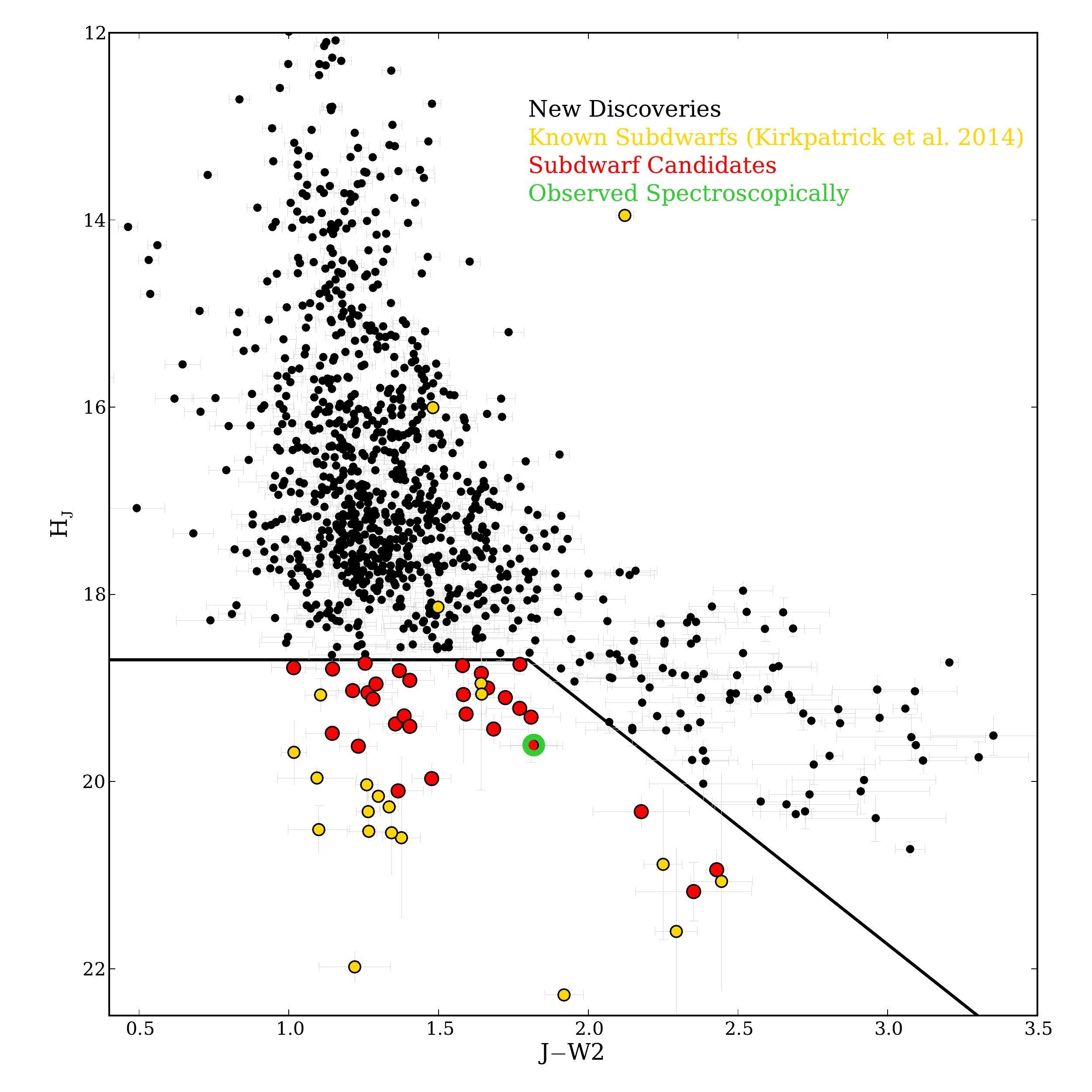}
\caption{Reduced proper motion diagram in J for discoveries from this survey.  Color coding is the same as in Figure 6.  The solid line denotes our subdwarf candidate selection criteria.}  
\end{figure}

\subsection{Comparison with \cite{kirk14} and \cite{luh14a} {\it WISE} motion Surveys} 

Figure 8 shows a comparison of the total proper motions and W2 magnitudes for all of the discoveries from this survey along with those from the {\it WISE} surveys of \cite{kirk14} and \cite{luh14a}.  While the AllWISE survey of \cite{kirk14} reported the largest number of new discoveries, they are by and large brighter and moving slower than those in the \cite{luh14a} and NEOWISE surveys.  As seen in the left panel of the figure, the majority of the discoveries from this survey have total proper motions between 250 and 400 mas yr$^{-1}$, a similar result to the \cite{luh14a} survey, while the majority of the discoveries from the \cite{kirk14} AllWISE motion survey have total motions less than $\sim$250 mas yr$^{-1}$, beyond the limit of our NEOWISE survey (see Section 2).  

\begin{figure*}
\plotone{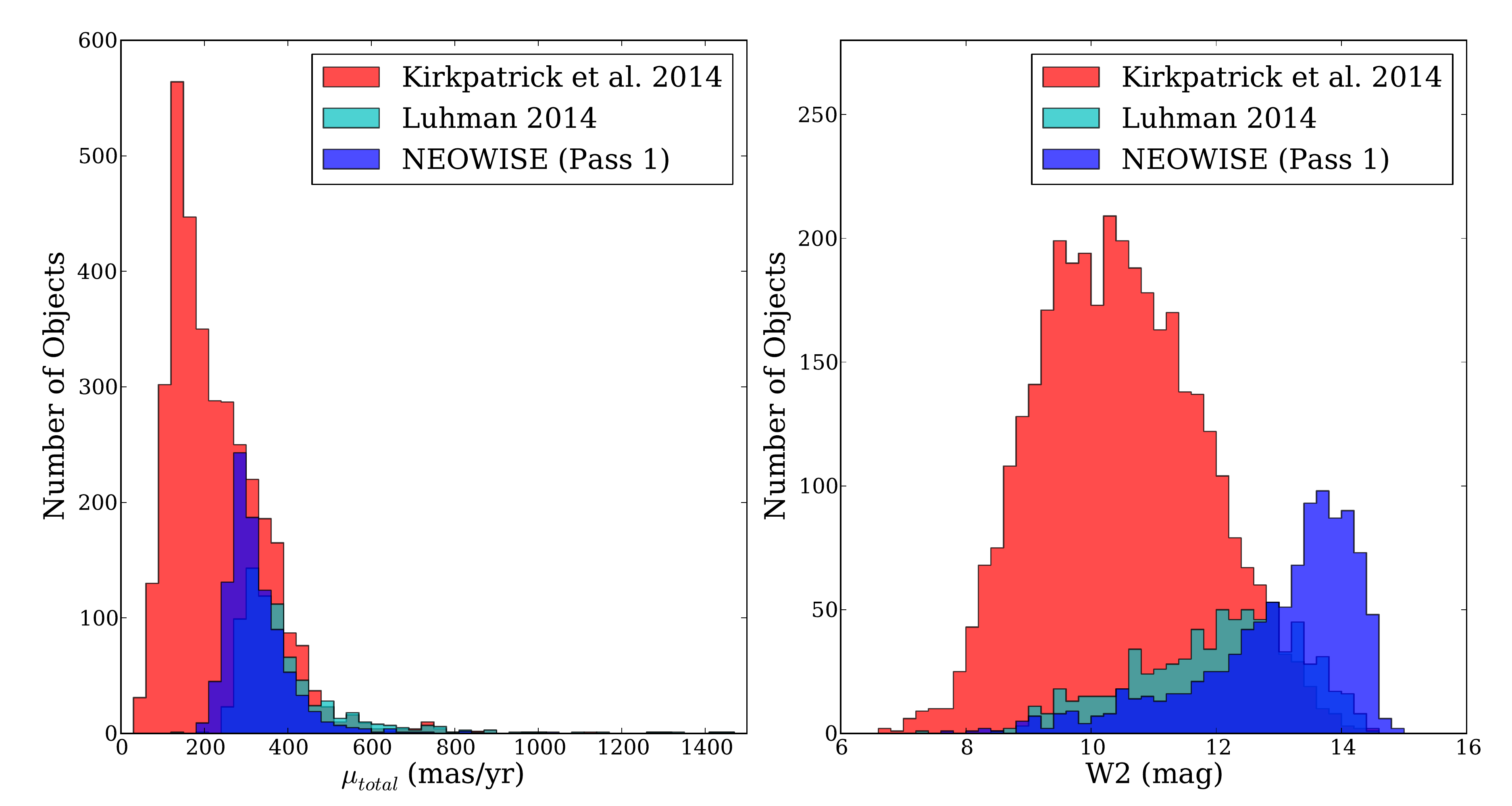}
\caption{The distribution of discoveries from the NEOWISE proper motion survey, the AllWISE motion survey \citep{kirk14}, and the {\it WISE} motion survey of \cite{luh14a} in both total proper motion and W2 magnitude. }  
\end{figure*}

As seen in the right panel of the figure, our NEOWISE motion survey has found significantly more objects at fainter magnitudes than the {\it WISE} surveys of \cite{kirk14} and \cite{luh14a}.  Figure 9 shows the same NEOWISE histogram from the right panel of Figure 8 broken up by estimated spectral type. As seen in the figure, almost every one of our new L and T dwarf candidates is contained within the fainter W2 magnitude bins. Of the 187 NEOWISE discoveries with estimated spectral types later than L0, 170 ($\sim$91\%) have W2 magnitudes greater than 13.  

\begin{figure}
\plotone{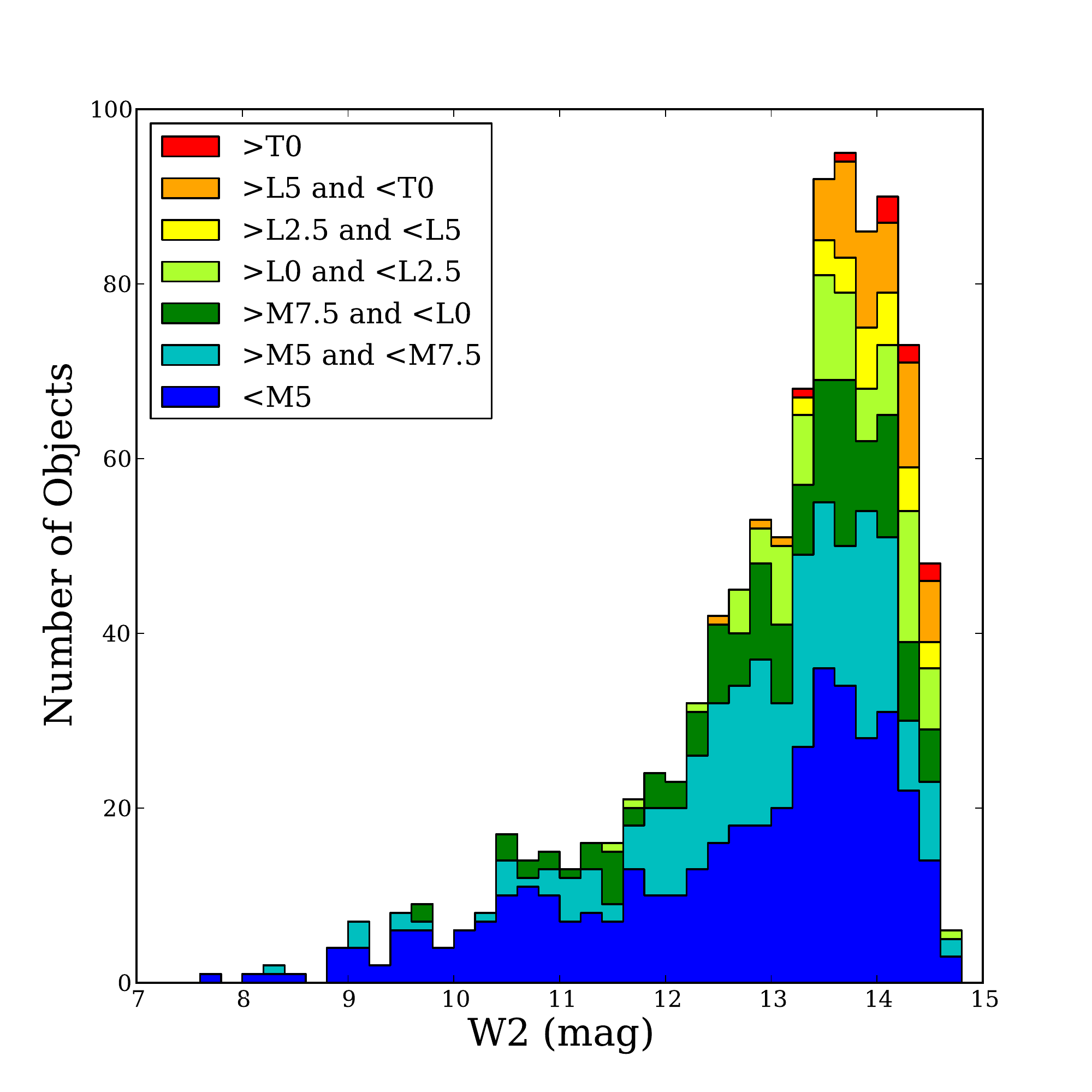}
\caption{The distribution of discoveries from our NEOWISE proper motion survey in W2 magnitude.  Colors correspond to photometrically estimated spectral types (see Section 3.3).}  
\end{figure}

The enhanced sensitivity of our survey to fainter high proper motion objects compared to the previous {\it WISE} motion surveys has allowed us to identify several new brown dwarfs with estimated spectral types around and later than the L/T transition.  Figure 10 shows a close-up view of the region of J$-$K$_{\rm S}$ vs.\ J$-$W2 color space occupied by L and T dwarfs.  In addition to the new discoveries from this survey, we also show the discoveries from the \cite{luh14a} and \cite{kirk14} surveys.  The figure shows that all of the surveys have identified several objects in the early to mid- L spectral type range, however only our NEOWISE survey identified new objects extending down into the mid-T dwarf color range.  Note that objects without 2MASS counterparts are not included in this figure.  Only one confirmed high proper motion object without a 2MASS counterpart was identified in the \cite{luh14a} and \cite{kirk14} surveys (WISE J085510.83$-$071442.5).  There are two such objects in our discovery list (see Sections 3.1 and 3.4).  All objects with estimated spectral types later than L7 are listed in Table 10.  We chose L7 to create this list because of the tendency of our classification program to mis-type early T-dwarfs as mid-Ls.  All 39 objects with estimated spectral types later than L7 are highlighted in Figure 10.  Follow-up spectroscopic observations for six of these objects (WISEA J001643.97$+$230426.5, WISEA J003338.45$+$282732.4, WISEA J010202.11$+$035541.4, WISEA J172120.69$+$464025.9, WISEA J223343.53$-$133140.9 and WISEA J230329.45$+$315022.7) are discussed in Section 5.2.  Spectral types for all of these objects are determined to be later than L7, with the exception of WISEA J003338.45$+$282732.4, a blue L3.   

\begin{deluxetable}{cc}
\tablecaption{New $>$L7 Brown Dwarf Candidates}
\tablehead{
\colhead{AllWISE } & Type\tablenotemark{a}  \\
Designation & (photometric) }
\startdata
J000430.66$-$260402.3 & 20.5\\
J000536.63$-$263311.8 & 17.1\\
J000856.39$-$281321.7 & 18.0\\
J001643.97$+$230426.5 & 18.9\\
J003338.45$+$282732.4 & 17.4\\
J010202.11$+$035541.4 & 18.9\\
J010631.20$-$231415.1 & 18.2\\
J013525.38$+$020518.2 & 17.7\\
J022721.93$+$235654.3 & 19.4\\
J024502.87$-$744519.3 & 17.2\\
J030119.39$-$231921.1 & 20.5\\
J031627.79$+$265027.5 & 19.0\\
J032309.12$-$590751.0 & 26.2\\
J032744.41$-$620336.3 & 17.7\\
J032838.73$+$015517.7 & 18.5\\
J034409.71$+$013641.5 & 19.1\\
J034858.75$-$562017.8 & 22.6\\
J041318.68$+$210326.5 & 17.7\\
J041743.13$+$241506.3 & 23.7\\
J051526.68$-$230954.2 & 17.1\\
J060202.67$+$724235.4 & 18.7\\
J062858.69$+$345249.2 & 17.2\\
J063552.52$+$514820.4 & 17.4\\
J084254.56$-$061023.7 & 22.7\\
J101944.62$-$391151.6 & 24.0\\
J103534.63$-$071148.2 & 17.7\\
J105131.36$-$144017.2 & 19.4\\
J135501.90$-$825838.9 & 17.1\\
J141127.86$-$481150.6 & 20.5\\
J172120.69$+$464025.9 & 18.2\\
J173551.56$-$820900.3 & 24.3\\
J192714.29$+$383754.2 & 17.0\\
J211157.84$-$521111.3 & 19.7\\
J211219.83$-$491717.0 & 17.8\\
J223343.53$-$133140.9 & 17.2\\
J223444.44$-$230916.1 & 17.4\\
J224931.10$-$162759.6 & 17.1\\
J230329.45$+$315022.7 & 18.3
\enddata
\tablenotetext{a}{Numerical spectral type estimates (e.g., 5 = M5, 15 = L5, 25 = T5)}
\end{deluxetable}

We can also place constraints on the existence of additional extremely cold, nearby WISE J085510.83$-$071442.5-type objects.  WISE J085510.83$-$071442.5 has a W2 magnitude of 14.02 $\pm$ 0.05 at a distance of 2.02 pc \citep{luh14d}.  Using our W2 survey limit 14.5, we can rule out the existence of additional J085510.83$-$071442.5-type objects with total proper motions between 0.25 and 15$\arcsec$ yr$^{-1}$ out to $\sim$2.9 pc.  Using the absolute magnitude$-$spectral type relations from \cite{dup12} and our W2 survey limit, we can also rule out the existence of additional Y0 and Y1 type dwarfs with proper motions between 0\farcs25 yr$^{-1}$ and 15$\arcsec$ yr$^{-1}$ out to distances of $\sim$11.5 pc and $\sim$9.5 pc, respectively.  A substantial increase in survey depth will be needed to place further constraints on the existence of such late-type objects in the Solar neighborhood.         

\begin{figure*}
\plotone{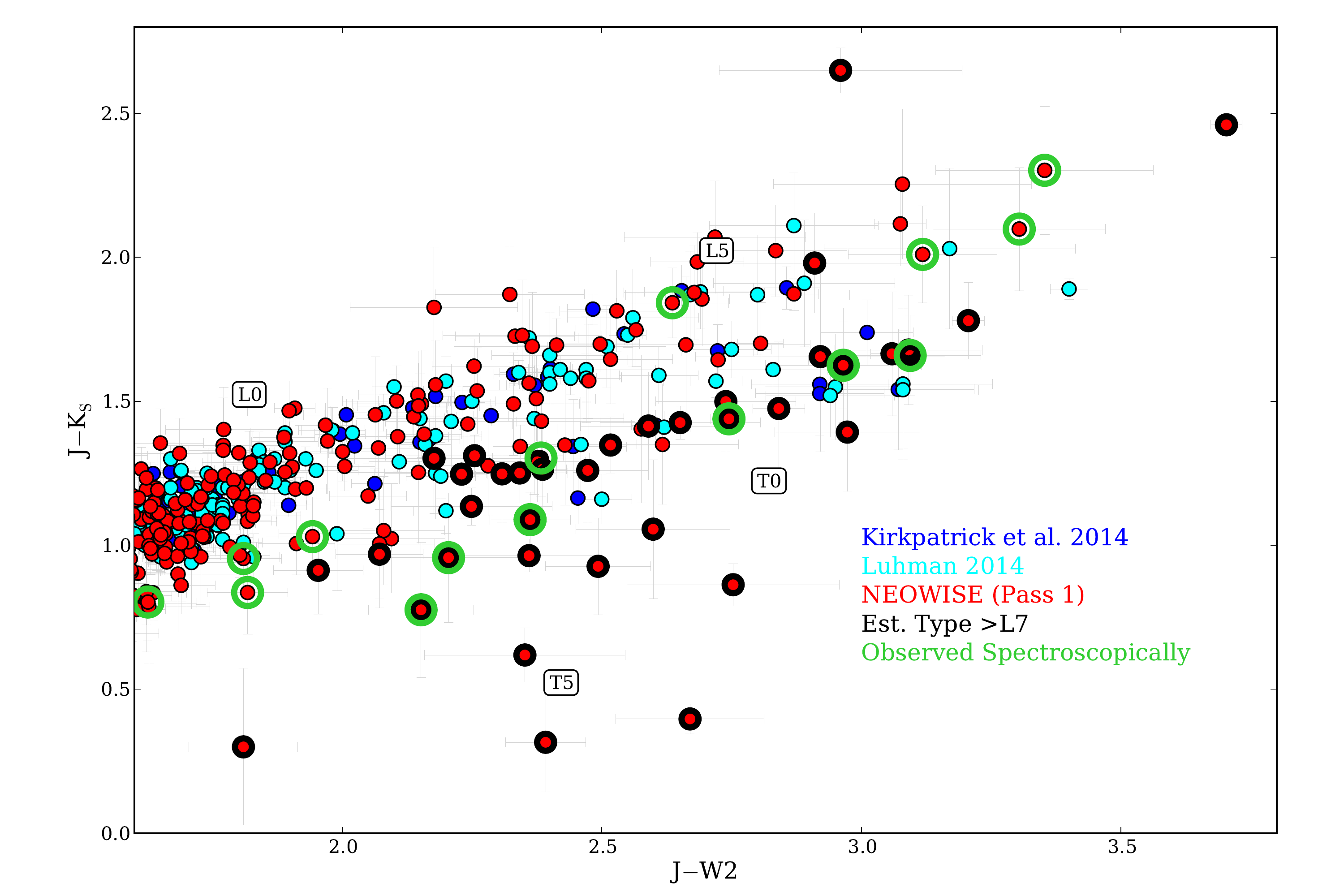}
\caption{J$-$K$_{\rm S}$ vs.\ J$-$W2 color-color diagram showing our new L and T type discoveries from \cite{luh14a}, \cite{kirk14}, and this survey.  Discoveries from this survey with estimated spectral types later than L7 are highlighted in black.}  
\end{figure*}

\section{Follow-up Observations}

\subsection{IRTF/SpeX}
Low resolution ($\lambda/\Delta\lambda$ = 75 $-$ 120) spectra were acquired for several sources with the upgraded SpeX spectrograph \citep{ray03} at the 3 m NASA Infrared Telescope Facility (IRTF) on Mauna Kea.  All observations were conducted using the prism mode.  A series of exposures was taken using an ABBA nod pattern along the 15$\arcsec$ long slit for each object.  A0V stars were observed at a similar airmasses for telluric correction purposes. Data for all objects were reduced using the SpeXtool reduction package (\citealt{cush04}; \citealt{vacca03}).  A summary of all IRTF/SpeX observations is given in Table 11.  

\subsection{Palomar/DoubleSpec}
Three targets were observed with the Double Spectrograph on the Hale 5m telescope on the night of UT 07 September 2015.  For the blue side of the spectrograph, we employed a 600 lines mm$^{-1}$ grating blazed at 4000 \AA\ for a total range of spectral coverage from 4015$-$7085 \AA.  For the red side of the spectrograph, we used a 600 lines mm$^{-1}$ grating blazed at 10,000 \AA\ for a total range of spectral coverage from 6545$-$9910 \AA. The overlapping regions were used to create one continuous spectrum across the entire range.  The flux standard used was Wolf 1346, which was bootstrapped to the flux calibration of \cite{ham94} using the standard Hiltner 600, both of which had been observed in an earlier run with the same setup on UT 27 Sep 2014.  Data were reduced using standard reduction procedures.

\begin{deluxetable*}{ccccccc}
\tablecaption{Observations}
\tablehead{
\colhead{AllWISE } & Sp.\ Type & Type\tablenotemark{a} & Dist. & $V_{\rm tan}$ & Obs. Date & Exp. Time\tablenotemark{b} \\
Designation &  & (photometric) & (pc) & (km s$^{-1}$) & (UT) & (s)}
\startdata
\cutinhead{IRTF/SpeX}
J000627.85$+$185728.8 & L7 & 16.1 & 31$-$35 & 43$-$50 &  27 June 2015 & 1200\\
J001643.97$+$230426.5 & T0 & 18.9 & 23$-$27 & 42$-$50 &  19 July 2015 & 1200\\
J003338.45$+$282732.4 & L3 (blue) & 17.4 & 38$-$44 & 59$-$69 &  27 June 2015 & 1200\\
J010202.11$+$035541.4 & L9 & 18.9 & 25$-$29 & 44$-$52 &  19 July 2015 & 1200\\
J011639.05$-$165420.5 & d/sdM8.5 & 15.8 & 78$-$92 & 213$-$252 &  27 June 2015 & 1200\\
J013012.66$-$104732.4 & d/sdM8.5 & 10.9 & 78$-$92 & 130$-$155 &  26 Feb 2015 & 1200\\
J114553.61$-$250657.1 & d/sdM7 &  8.4 & 107$-$128 & 129$-$156 &  9 May 2015 & 1200\\
J120035.40$-$283657.5 & T0 & 16.7 & 22$-$26 & 56$-$58 &  28 Jan 2015 & 960\\
J122221.95$-$213948.6 & L6 & 14.4 & 28$-$33 & 45$-$54 &  8 May 2015 & 1200\\
J130729.56$-$055815.4 & L8 (sl. blue) & 16.9 & 26$-$30 & 45$-$53 &  28 Jan 2015 & 1440\\
J144033.28$-$080406.9 & L2 (blue) & 15.2 & 50$-$58 & 78$-$92 &  9 May 2015 & 1200\\
J170726.69$+$545109.3 & L1 (blue) & 13.0 & 59$-$68 & 84$-$97 &  27 June 2015 & 1200\\
J172120.69$+$464025.9 & T0: (pec) & 18.2 & 26$-$30 & 33$-$39 &  27 June 2015 & 1200\\
J205202.06$-$204313.0 & L8 (sl. blue) & 15.9 & 27$-$31 & 50$-$58 &  27 June 2015 & 1200\\
J223343.53$-$133140.9 & T2 (blue) & 17.2 & 26$-$30 & 39$-$46 &  19 July 2015 & 1200\\
J230329.45$+$315022.7 & T2 (blue) & 18.3 & 24$-$28 & 32$-$38 &  27 June 2015 & 1200\\
\cutinhead{Palomar/DoubleSpec}
J172602.92$-$034211.7 & sdM1.5 & 5.3 & 95$-$151 & 126$-$201 & 1 September 2015 & 2400,2460 \\
J221126.37$-$192207.4 & M5/sdM6 & 10.2 & 65$-$99 & 100$-$153 & 1 September 2015 & 2400,2460 \\
J232656.09$-$181504.5 & M5 & 9.5 & 62$-$97 & 89$-$137 & 1 September 2015 & 2400,2460 
\enddata
\tablenotetext{a}{Numerical spectral type estimates (e.g., 5 = M5, 15 = L5, 25 = T5)}
\tablenotetext{b}{The two exposure times listed for Palomar/DoubleSpec observations refer to the blue and red sides of the spectrograph.}
\end{deluxetable*}

\section{Analysis}
\subsection{Spectral Classification}
We determine spectral types for all near-infrared spectra following the method outlined in the Appendix of \cite{schneid14}.  Comparisons of each acquired spectrum with its best matching near-infrared spectral standard from the Spex Prism Spectral Library \citep{burg14}\footnote{http://pono.ucsd.edu/$\sim$adam/browndwarfs/spexprism/library.html} are shown in Figure 11.  Spectral types are provided in Table 11.  We also include the spectral type estimates from Table 1 for comparison.  It should come as no surprise that the estimated spectral types differ significantly from the actual spectral types for these objects because many are poor matches to the spectral standards and occupy unique regions of color space (see Figures 5, 6, 10, and 11 and Section 4.4).  The three L dwarfs that match well at all near-infrared wavelengths with their corresponding spectral standard (WISEA J122221.95$-$213948.6 , WISEA J000627.85$+$185728.8, and WISEA J010202.11$+$035541.4) all have photometric spectral type estimates within $\sim$1.5 subtypes of their actual type.  While the two new T0 dwarfs (WISEA J001643.97$+$230426.5 and WISE J120035.40$-$283657.5) are good matches to the T0 standard, their estimated types are several subtypes earlier (18.9 and 16.7, respectively).  This is not unexpected, as early T dwarfs typically have photometric spectral type estimates earlier than their actual type using our classification method (see Appendix A).  We estimate the distance to each observed object using the spectral types determined from the comparison with spectral standards, W2 magnitudes, and the absolute magnitude$-$spectral type relations from \cite{dup12} and provide distance ranges in Table 11 using a $\pm$0.5 subtype uncertainty and photometric uncertainties.         

All optical spectra were classified based on the classification system of \cite{kirk91} for normal M dwarfs or the subdwarf classification scheme of \cite{lep07a}.  

\subsection{Individual Objects of Note}

{\it WISEA J001643.97$+$230426.5 and WISEA J010202.11$+$035541.4:} Both WISEA J001643.97$+$230426.5 and WISEA J010202.11$+$035541.4 were singled out as potentially nearby (Table 7) and late-type (Table 10).  Both objects have photometric type estimates similar to their actual spectral types (see Table 9), and therefore have similar spectral type distance estimates.  Both of these brown dwarfs are estimated to be within $\sim$30 pc.  

{\it WISEA J011639.05$-$165420.5, WISEA J013012.66$-$104732.4, and WISEA J114553.61$-$250657.1:} All three of these objects were chosen for follow-up because they have colors indicative of being late M-type subdwarfs (see Figure 6 and Table 8).  WISEA J011639.05$-$165420.5 was also chosen as a subdwarf candidate from its placement on the reduced proper motion diagram in Figure 7. While each of these three objects match reasonably well to either the M7 or M8 near-infrared spectral standard in the J band, they are all poor matches at H and K because they are much bluer than the standards.  This is a characteristic typical of late-type subdwarfs.  In Figure 12 we show the near-infrared spectra of these three objects compared with subdwarf spectra from the Spex Prism Spectral Library \citep{burg14}.  As seen in the figure, WISEA J114553.61$-$250657.1 closely resembles 2MASS J18355309$-$3217129, which is classified as d/sdM7 in \cite{kirk10}.   The figure also shows that both WISEA J011639.05$-$165420.5 and WISEA J013012.66$-$104732.4 are similar to LSR 1826$+$3014, classified as d/sdM8.5 in \cite{burg04}.  While the relations from \cite{dup12} are designed for normal objects, not subdwarfs, we still use the relations these three objects only as a preliminary distance estimates.  Given their new spectral classifications, we estimate distance ranges for WISEA J011639.05$-$165420.5, WISEA J013012.66$-$104732.4, and WISEA J114553.61$-$250657.1 of 78$-$92, 78$-$92, and 107$-$128 pc, respectively.  Using these distance estimates, along with their proper motions, we consider whether any of these objects are part of the thick disk/halo population based on their tangential velocities ($V_{\rm tan}$).  We find $V_{\rm tan}$ ranges of 213$-$252, 130$-$155, and 129$-$156 km s$^{-1}$, which do indeed point towards membership in the thick disk/halo population, as \cite{fah09} find a median tangential velocity of 26 km s$^{-1}$ with a dispersion of 19 km s$^{-1}$ for late-M dwarfs.  Membership in the thick disk/halo is not surprising given their blue near-infrared colors and subdwarf spectral classifications.  Future optical spectroscopy could confirm each of these object's low metallicity and their subdwarf classification.       

{\it WISEA J003338.45$+$282732.4, WISEA J144033.28$-$080406.9, and WISEA J170726.69$+$545109.3:} WISEA J003338.45$+$282732.4 was identified as a potentially nearby object (Table 7) and as a potential T dwarf (Table 10).  All three of these objects match well to early L spectral standards at J, but are much bluer overall.  Surface gravity and/or low metallicity are thought to account for the blue color of blue L dwarfs, which is supported by their kinematics \citep{fah09}.  We classify each of these three objects as early type blue L dwarfs.  We calculate $V_{\rm tan}$ ranges of 59$-$69, 78$-$92, and 84$-$97 km s$^{-1}$ for WISEA J003338.45$+$282732.4, WISEA J144033.28$-$080406.9, and WISEA J170726.69$+$545109.3, respectively, using their photometric distance estimates.  These values are higher than the median tangential velocities for L dwarfs of $\sim$30 km s$^{-1}$ found in previous studies (\citealt{schm10}, \citealt{fah12}).  Optical spectroscopy would be able to determine if the blue nature of these L dwarfs is due to low-metallicity.

{\it WISEA J130729.56$-$055815.4 and WISEA J205202.06$-$204313.0:}  Both of these objects are slightly bluer  than the L8 spectral standard.  We classify each as L8 (sl. blue).     

{\it WISEA J172120.69$+$464025.9:}  WISEA J172120.69$+$464025.9 was selected as a potential late-type brown dwarf (Table 10).  This object does not match well with any of the spectral standards, but generally displays the overall shape of a T0.  We classify this object as T0: (pec).  We investigated spectral binarity as a possible explanation for this object's unusual spectrum (e.g., \citealt{burg07a}, \citealt{bar14}), but could not find a satisfactory fit.    

{\it WISEA J223343.53$-$133140.9 and WISEA J230329.45$+$315022.7:}  WISEA J223343.53$-$133140.9 and WISEA J230329.45$+$315022.7 were both selected as potential late-type brown dwarfs (Table 10).  Both of these objects are excellent matches to the T2 spectral standard at J, however both are significantly bluer than the standards.  We therefore classify them at T2 (blue).  

{\it WISEA J172602.92$-$034211.7}:  This object was chosen for follow-up spectroscopy because it stood out prominently in color space, with $J-K_S$ and $J-W2$ values of 1.13 and 1.17 mag, respectively (see Figure 6).  Optical spectroscopy revealed this object to be a metal-poor early M-type star (Figure 13).  Comparison with the sdM standards of \cite{lep07a} show good agreement with the sdM1 and sdM2 standards.  We therefore classify WISEA J172602.92$-$034211.7 as an sdM1.5. 

{\it WISEA J221126.37$-$192207.4 and WISEA J232656.09$-$181504.5}:  These objects are color-selected subdwarf candidates (see Table 8).  WISEA J232656.09$-$181504.5 may be slightly metal-poor, but is overall a good match to the normal M5 standard (Figure 13) and therefore classified as M5.  Figure 13 also shows that WISEA J221126.37$-$192207.4 matches fairly well with the M5 standard.  However, this object's spectrum does show slightly enhanced CaH absorption around 7000 \AA, which is typical of M subdwarfs \citep{lep07a}. We also show in Figure 13 a comparison with the sdM6 standard, which matches fairly well.  We measure a $\zeta$$_{\rm TiO/CaH}$ metallicity index of 0.865, slightly above the cutoff value between normal dwarfs and subdwarfs of 0.825 given in \cite{lep07a}. We conservatively give WISEA J221126.37$-$192207.4 a spectral type of M5/sdM6.      

\begin{figure*}
\includegraphics[scale=0.60,angle=90]{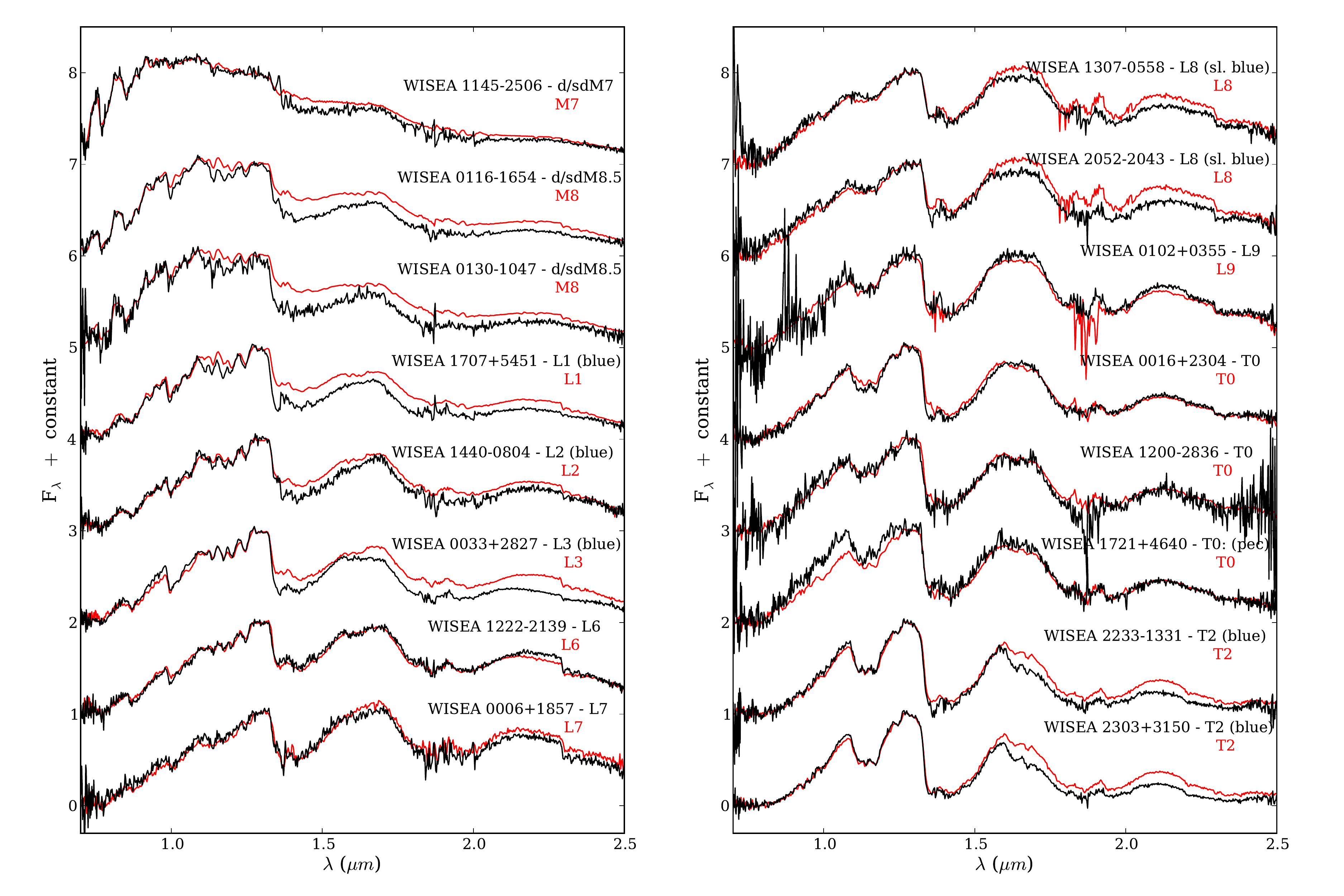}
\caption{IRTF/SpeX spectra of new M, L, and T dwarfs compared to near-infrared spectral standards (red).  All spectra are normalized at 1.28 $\mu$m.  The spectral standards are: VB 8 (M7; \citealt{burg08}) VB 10 (M8; \citealt{burg04}), 2MASSW J2130446$-$084520 (L1; \citealt{kirk10}), Kelu$-$1 (L2; \citealt{burg07b}), 2MASSW J1506544$+$132106 (L3; \citealt{burg07b}), 2MASSI J1010148$-$040649 (L6; \citealt{reid06}), 2MASSI J0103320$+$193536 (L7; \citealt{cruz04}), 2MASSW J1632291$+$190441 (L8; \citealt{burg07b}), DENIS-P J0255$-$4700 (L9; \citealt{burg06a}), SDSS J120747.17$+$024424.8 (T0; \citealt{loop07}) and SDSSp J125453.90$-$012247.4 (T2; \citealt{burg04}).}  
\end{figure*}

\begin{figure}
\plotone{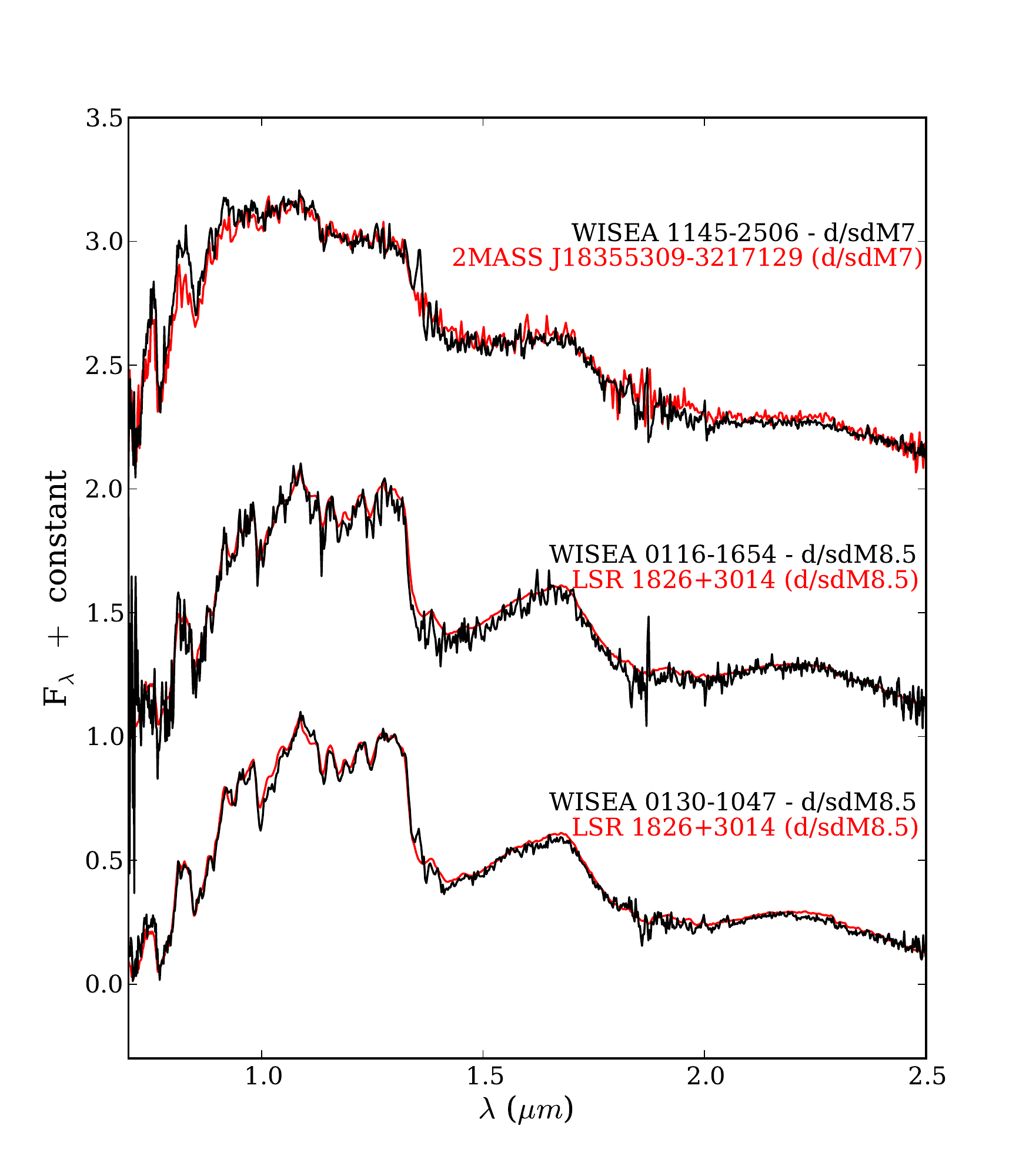}
\caption{IRTF/SpeX spectra of 3 late-M subdwarf candidates.  Comparison spectra are 2MASS J18355309$-$3217129 (d/sdM7 - \citealt{kirk10}) and LSR 1826$+$3014 (d/sdM8.5 - \citealt{burg04}).}  
\end{figure}

\begin{figure*}
\plotone{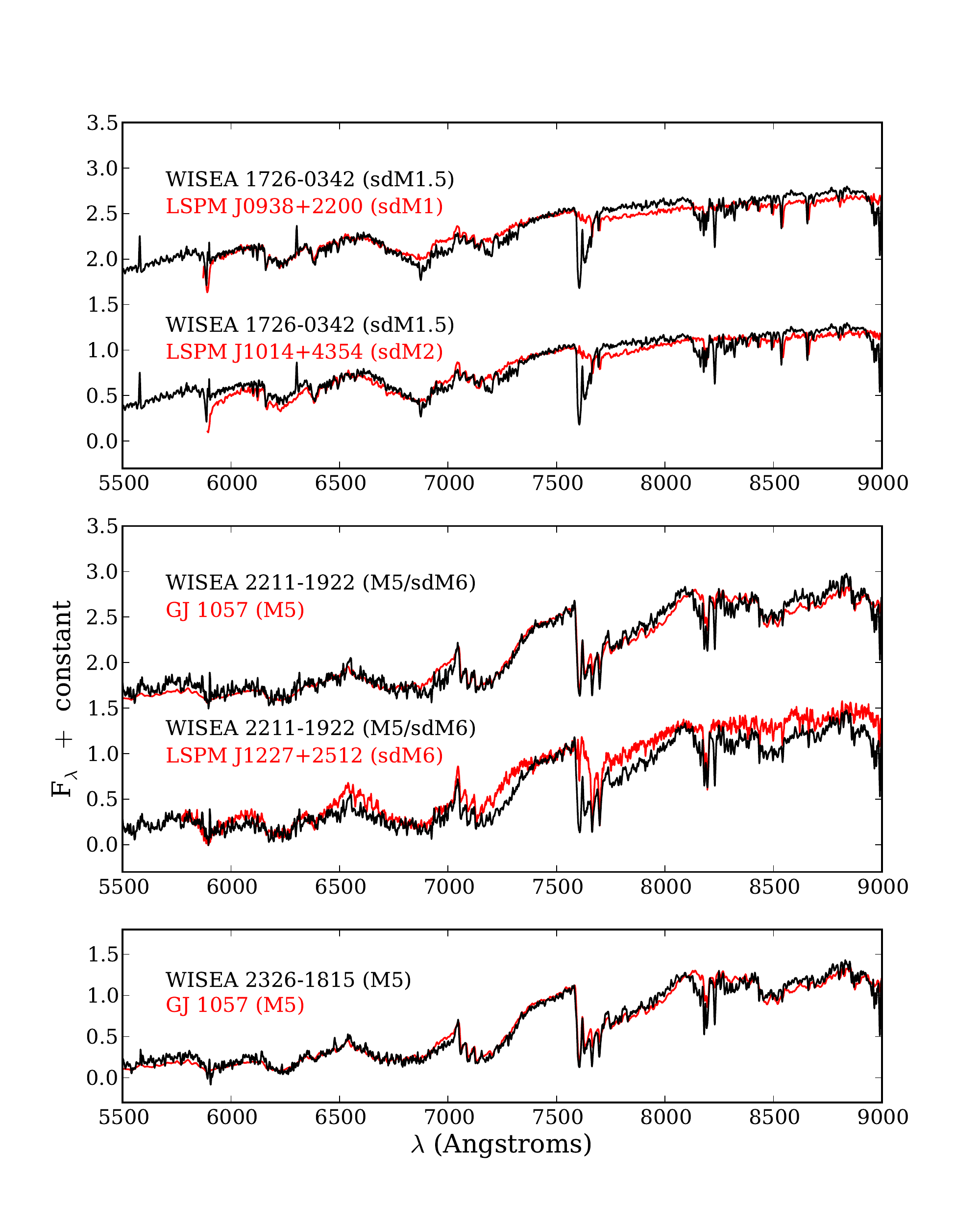}
\caption{Palomar/DoubleSpec optical spectra (black) compared to dwarf and subdwarf spectral standards (red).  All spectra are normalized at 7500 \AA.  The spectral standards are: LSPM J0938$+$2200 (sdM1; \citealt{lep07a}), LSPM J1014$+$4354 (sdM2; \citealt{lep07a}), GJ 1057 (M5; \citealt{kirk97}), and LSPM J1227$+$2512 (sdM7; \citealt{lep07a}). Note that the subdwarf standards from \cite{lep07a} are corrected for telluric absorption and our Palomar DoubleSpec spectra are not.  }  
\end{figure*}

\section{Conclusion}
We have conducted a survey for high proper motion objects using the first sky pass of NEOWISE and the AllWISE catalog, identifying over twenty thousand high proper motion objects, over one thousand of which are new discoveries. Through an analysis of 2MASS and AllWISE colors and estimated spectral types, we have picked out a number of appealing candidates identified as being nearby objects, subdwarfs, or late-type brown dwarfs, several of which have been confirmed with near-infrared or optical spectroscopy.  The success of this survey, and the previous motion surveys of \cite{luh14a} and \cite{kirk14}, demonstrates the effectiveness of using data from the {\it WISE} telescope to identify previously overlooked objects of scientific interest. The foremost limiting factor for these surveys has been the depth at which objects with large motions can be readily identified.  A future catalog produced from coadding the NEOWISE single frames would significantly increase the survey volume of this type of effort, and would only enlarge the already substantial legacy of the {\it WISE} telescope. 

\acknowledgments

The authors wish to thank S{\'e}bastian L{\'e}pine for graciously providing us with subdwarf standard spectra.  This publication makes use of data products from the {\it Wide-field Infrared Survey Explorer}, which is a joint project of the University of California, Los Angeles, and the Jet Propulsion Laboratory/California Institute of Technology, and NEOWISE, which is a project of the Jet Propulsion Laboratory/California Institute of Technology. {\it WISE} and NEOWISE are funded by the National Aeronautics and Space Administration.  This research has benefitted from the M, L, T, and Y dwarf compendium housed at DwarfArchives.org.  This research has made use of the VizieR catalog access tool and SIMBAD database operated at, CDS, Strasbourg, France.  Funding for SDSS-III has been provided by the Alfred P. Sloan Foundation, the Participating Institutions, the National Science Foundation, and the U.S. Department of Energy Office of Science. The SDSS-III web site is http://www.sdss3.org/.  SDSS-III is managed by the Astrophysical Research Consortium for the Participating Institutions of the SDSS-III Collaboration including the University of Arizona, the Brazilian Participation Group, Brookhaven National Laboratory, Carnegie Mellon University, University of Florida, the French Participation Group, the German Participation Group, Harvard University, the Instituto de Astrofisica de Canarias, the Michigan State/Notre Dame/JINA Participation Group, Johns Hopkins University, Lawrence Berkeley National Laboratory, Max Planck Institute for Astrophysics, Max Planck Institute for Extraterrestrial Physics, New Mexico State University, New York University, Ohio State University, Pennsylvania State University, University of Portsmouth, Princeton University, the Spanish Participation Group, University of Tokyo, University of Utah, Vanderbilt University, University of Virginia, University of Washington, and Yale University.

\begin{appendix}
\section{Photometric Spectral Type Estimates}
In order to prioritize the most interesting objects from our list of new discoveries for follow-up observations, we endeavored to find a way to accurately estimate approximate spectral types for each object using solely 2MASS and AllWISE photometry.  While previous studies have attempted photometric typing, mainly using color-spectral type polynomial relations (e.g., \citealt{luh14c} and \citealt{skr15}), machine learning algorithms are an alternative tool to use to accomplish this type of classification.  For this work, we utilized k-Nearest Neighbors (k-NN) algorithm using code available from the scikit-learn project \citep{ped12}.  The k-NN algorithm classifies by identifying the closest training data points within the space being examined.  For a test sample, the Euclidean distance is calculated for each member of the comparison data set.  The k value determines how many training data points are selected.  The test sample is then classified into the training set that is most common amongst its k nearest neighbors.  This requires a well defined training set of known objects with known spectral types.  We used an updated list of M, L, and T dwarfs from DwarfArchives.org with near-infrared spectral types (C. Gelino, priv. comm.).  For each object from the DwarfArchives list, we found its corresponding 2MASS and AllWISE catalog entries, retaining only those objects that had both.  In order to ensure that the estimated spectral classifications are not biased towards spectral classes for which there is a larger population of objects, we limit the training set to have a maximum of 10 objects per half spectral type bin.  Objects included in the training set were preferentially chosen to have the smallest photometric uncertainties.  

We evaluate the accuracy of photometrically classifying objects using the k-NN algorithm using two different test sets.  For the first, we randomly selected 10\% of objects from the training set.  For the second, we use the entire list of M, L, and T dwarfs from the DwarfArchives list.  We then evaluate for each object in each test sample the probability of belonging to every spectral class for every possible color-color combination using 2MASS J, H, and K$_S$ and AllWISE W1 and W2 magnitudes (45 total).  A final spectral type estimate for each object in the test sample is determined by summing the product of the probabilities for each spectral type and the numbered index for that spectral type (e.g., M5 = 5, L5 = 15, T5 = 25).  We repeat the procedure for the first test set 1000 times, and test the accuracy by using two different metrics; the RMS and the median of the absolute differences (MAD), which we use in an attempt to account for outliers, defined as:

\begin{equation}
median \mid SpT_{actual} - SpT_{estimated} \mid
\end{equation}        
             
\noindent where $SpT_{actual}$ is the near-infrared spectral type from DwarfArchives, and $SpT_{estimated}$ is the spectral type determined by the algorithm.  A comparison of the estimated spectral types versus the actual spectral types for one run of the 10\% sample is shown in the top left panel of Figure 14.  The average and standard deviation of RMS and MAD values for the entire simulation of 10\% test samples are 1.14 $\pm$ 0.15  and 0.67 $\pm$ 0.10 subtypes, respectively.  However, the RMS and MAD values are spectral-type dependent, as shown in the right panel of Figure 14.  Almost all objects have estimated spectral types within 1.5 subtypes of their actual type.  Early T dwarfs (T2s and T3s) are consistently classified as several subtypes earlier than their actual type.  We suspect this is because mid-L dwarfs and early T dwarfs share similar near- and mid- infrared colors \citep{kirk11}.  For the entire list of M, L, and T dwarfs from DwarfArchives, we find slightly larger RMS and MAD values of 1.75 and 0.87, respectively.  A comparison of the estimated spectral types versus the actual spectral types for this entire sample is shown in the bottom left panel of Figure 14.  We see large RMS values for very early L dwarfs, approaching values as high as $\sim$3 subtypes for L0. This is clearly due to an excess outliers, as the MAD values are not so extreme.  These outliers may be actual photometric outliers, or potentially mistyped L dwarfs.  We see the same peak around T3 as seen in the 10\% sample, most likely for the same reasons.  For the vast majority of objects in the entire DwarfArchive near-infrared spectral type catalog, spectral types are accurate to within $\pm$2 subtypes.

\begin{figure}
\plotone{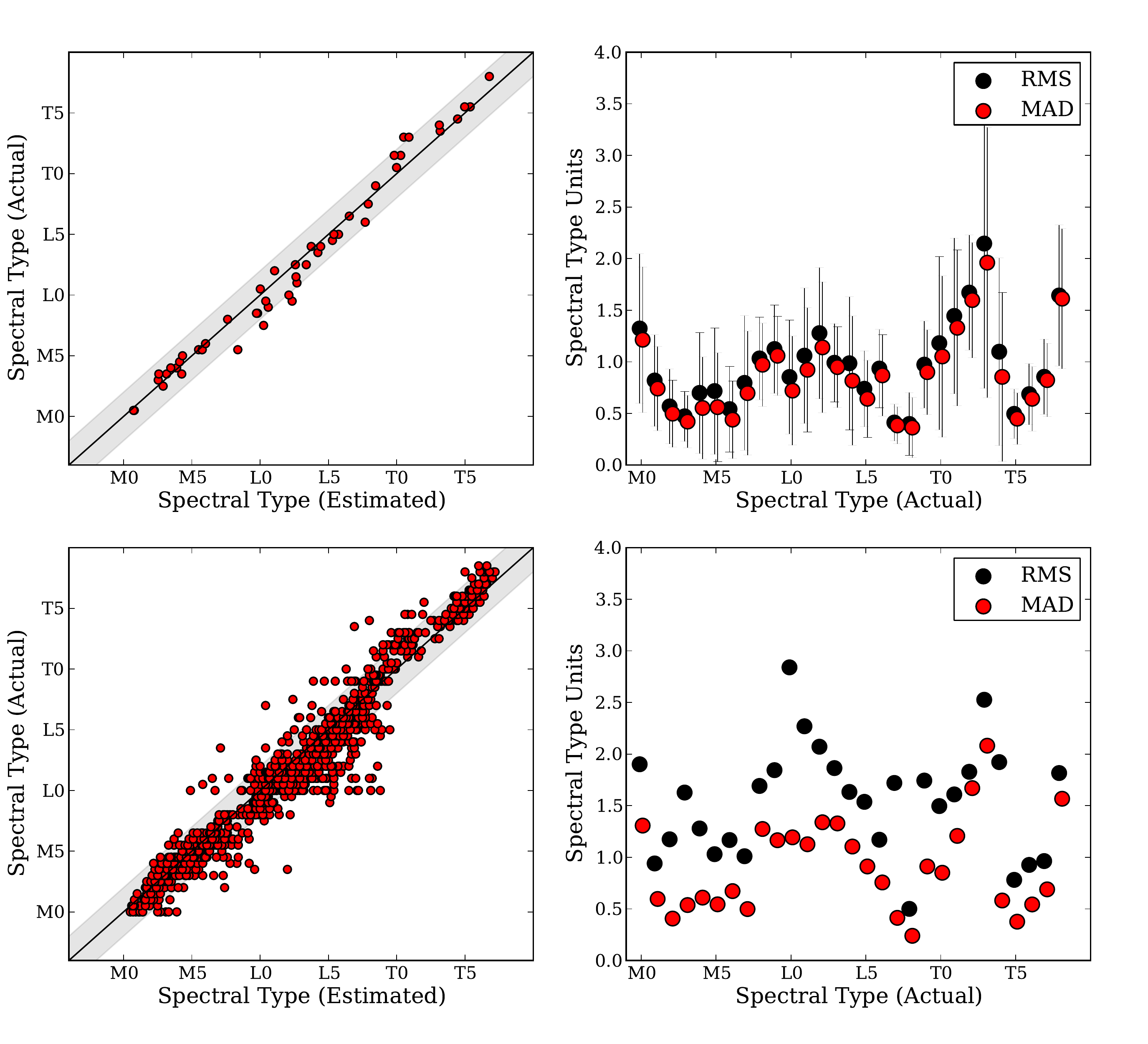}
\caption{{\it Top Left:} The actual versus estimated spectral types for a test sample comprising of 10\% of the training sample using the k-Nearest Neighbors classification algorithm.  The gray bar indicates $\pm$2 subtypes. {\it Top Right:} The RMS and the median of absolute differences (MAD - defined in equation A1) as a function of spectral type for the 1000 simulations . {\it Bottom Left:} The actual versus estimated spectral types for every M, L, and T dwarf with a near-infrared spectral type from DwarfArchives.  The gray bar indicates $\pm$2 subtypes. {\it Bottom Right:} The RMS and the MAD as a function of spectral type for the entire M, L, and T dwarf sample from DwarfArchives. }  
\end{figure}

To compare our results with those that use polynomial relations, we evaluate our classifications using the same robust estimator as that used in \cite{skr15}, namely

\begin{equation}
\sigma = \frac{\sum\limits_{i=1}^N \abs{\Delta t}}{N} \frac{\sqrt{2\pi}}{2}
\end{equation}        
       
\noindent For the entire DwarfArchive list, we find $\sigma_{\rm M}$ = 1.2, $\sigma_{\rm L}$ = 1.8, and $\sigma_{\rm T}$ = 1.4.  These values are similar to those found for the polynomial relations in \cite{skr15} of $\sigma_{\rm L}$ = 1.5 and $\sigma_{\rm T}$ = 1.2.  Note however that our method  uses only JHKW1W2 photometry, while the method in \cite{skr15} uses izYJHKW1W2, when available.

\end{appendix}


\end{document}